\documentclass[reprint,
 amsmath,amssymb,
 aps
]{revtex4-2}
\usepackage{bm, color}

\usepackage[colorlinks = true,
            allcolors = blue]{hyperref}
\usepackage{placeins}
\usepackage{float}
\usepackage{hyperref}

\usepackage{graphicx}
\usepackage{dcolumn}
\usepackage{bm}
\usepackage{lipsum}

\usepackage{color,soul}

\begin{document}

\preprint{APS/123-QED}

\title{A Real-time Dyson Expansion Scheme: Efficient Inclusion of Dynamical Correlations in Non-equilibrium Spectral Properties}
\author{Cian C. Reeves}
\affiliation{%
Department of Physics, University of California, Santa Barbara, Santa Barbara, CA 93117
}%

\author{Vojt\ifmmode \check{e}\else \v{e}ch Vl\ifmmode \check{c}\else \v{c}ek}
\affiliation{%
Department of Chemistry and Biochemistry, University of California, Santa Barbara, Santa Barbara, CA 93117
}%
\affiliation{%
Department of Materials, University of California, Santa Barbara, Santa Barbara, CA 93117
}

\date{\today}

\begin{abstract}
Time-resolved photoemission spectroscopy is the key technique to probe the real-time non-equilibrium dynamics of electronic states.  Theoretical predictions of the time dependent spectral function for realistic systems is however, a challenge.  Employing the Kadanoff-Baym equations to find this quantity results in a cubic scaling in the total number of time steps, quickly becoming prohibitive and often fail quantitatively and even qualitatively. In comparison, mean-field methods have more favorable numerical scaling both in the number of time steps and in the complexity associated with the cost of evolving for a single time step, however they miss key spectral properties such as emergent spectral features.  Here we present a scheme that allows for the inclusion of dynamical correlations to the spectral function while maintaining the same scaling in the number of time steps as for mean-field approaches,  while capturing the emergent physics. Further, the scheme can be efficiently implemented on top of equilibrium real-time many-body perturbation theory schemes and codes.  We see excellent agreement with exact results for test systems.  Furthermore we exemplify the method on a periodic system and demonstrate clear evidence that our proposed scheme produces complex spectral features including excitonic band replicas, features that are not observed using static mean-field approaches.
\end{abstract}

\maketitle
\paragraph*{Introduction}
In recent years there has been great interest in the non-equilibrium properties of quantum materials. These systems can display phenomena such as ultrafast  phase transitions\cite{Beebe_2017,Disa_2023}, formation of exotic states\cite{Dong_2021,Bao_2022} and allow tuning of material properties through driving\cite{Nuske_2020,Zhou_2023}.  The first principles description of non-equilibrium dynamics of electrons, motivated by advances in experiments \cite{Sie_2019,Karni_2023,Man_2021}, has attracted attention in recent years \cite{Perfetto_2022,Sun_2021,Karlsson_2021} and stimulated new theoretical developments \cite{Schlunzen_2020,Kaye_2021,Lian_2018}.  In particular the Hartree-Fock Generalized Kadanoff-Baym ansatz (HF-GKBA) emerged as a powerful tool\cite{Lipavsky_1986,Schlunzen_2020,Perfetto_2022}.  It has been shown in many scenarios to provide results that are as accurate as the Kadanoff-Baym equations (KBEs)\cite{Reeves_2023,Schlunzen_2017,Bonitz_2013,Balzer_2013,Hermanns_2013}, which HF-GKBA approximates. Unlike the KBEs,  the HF-GKBA does not require computation of the Green's function (GF) on the full $t,t'$ grid (Fig 1)\cite{Schlunzen_2020,Joost_2020}. Further, KBEs can suffer from spurious damping\cite{von_Friesen_2009,von_Friesen_2010}.  There is however one major caveat: these excellent results are restricted to properties derived from the time-diagonal component of the Green's function.  Since the HF-GKBA reconstructs the time off-diagonal components at the Hartree-Fock (HF) level, properties linked to these components perform similarly to HF results\cite{Tuovinen_2020}. In some cases the HF-GKBA results are even worse than HF alone\cite{supp}.  

Time off-diagonal components are however, the key ingredient in computing the non-equilibrium time-resolved spectral function\cite{Freericks_2009}. This gives information about the non-equilibrium energy distribution of quasiparticles in the system, and captures exciton formation and non-equilibrium phase transitions\cite{Boschini_2024,choi_2023,Disa_2023}.  
Furthermore, to see the intra-gap electron-hole signatures it is necessary to include two-body correlations when computing the time off-diagonals.  A feature which we will show is missed by HF and the HF-GKBA.

\begin{figure}
    \centering
    \includegraphics[width=.4\textwidth]{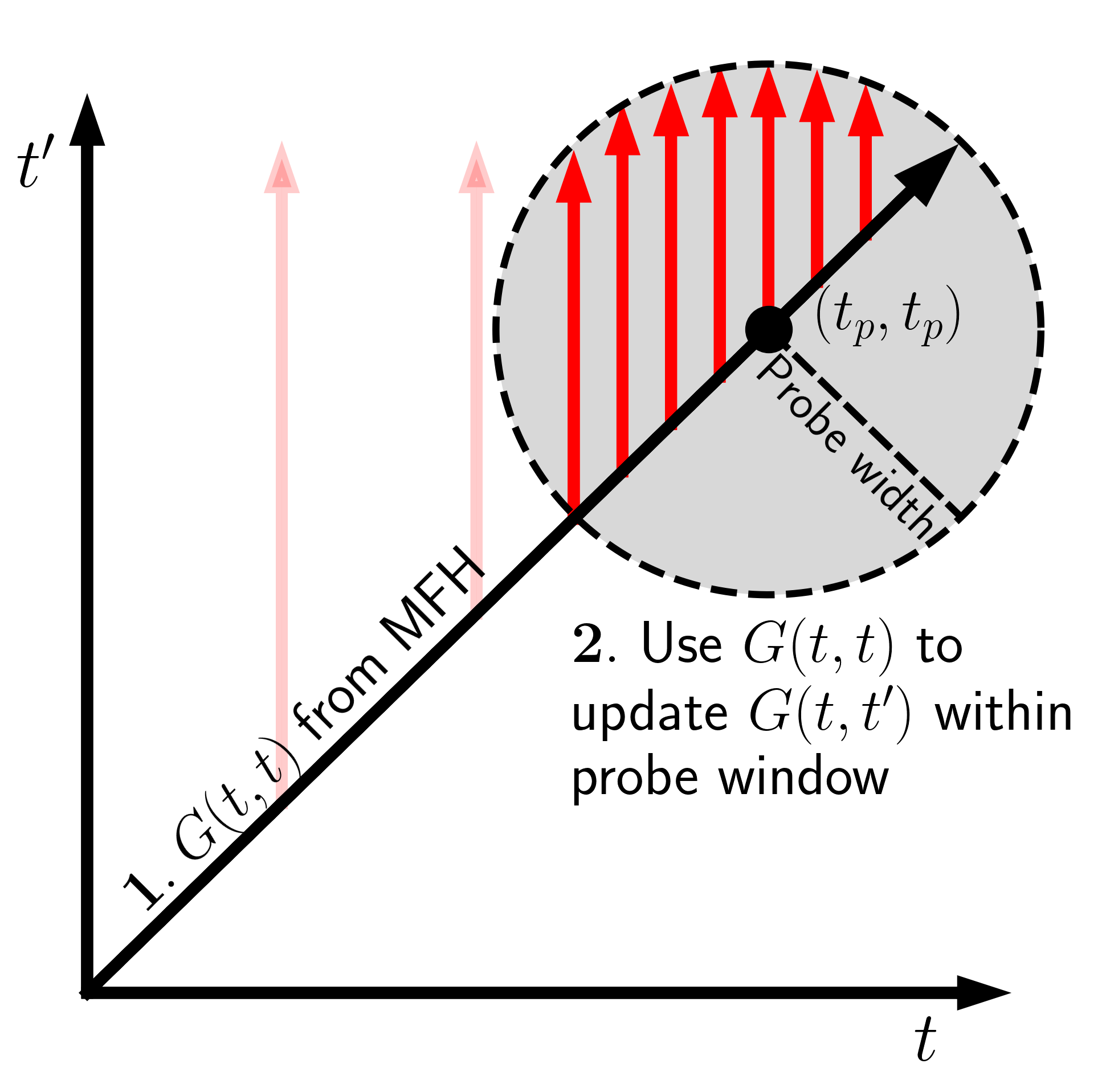}
    \caption{Schematic of the off-diagonal GKBA scheme.  In the two step procedure one first performs a time evolution along the time-diagonal.  The information from the diagonal is then used as a starting point to update the off-diagonal}
    \label{fig:RT-DE_schematic}
\end{figure}
While methods have been employed to improve the HF-GKBA, they have mainly been applied to improve the time-diagonal and steady-state properties\cite{Latini_2014,Hopjan_2019}. Works that seek to improve the spectral properties\cite{Schuler_2020} are scarce and the improvements do not go beyond the mean-field level due to the underlying reconstruction of the off-diagonal. Performing the full KBE calculations\cite{Stan_2009,stefanucci2013nonequilibrium} instead is however problematic as the computational scaling  is cubic in the number of time-steps, making it numerically impractical. Although methods exist to reduce the scaling of the KBEs\cite{Kaye_2021,Yin_2022,Kaye_2023,Stahl_2022,Blommel_2024}, they can also suffer from damping effects which leads to broadening of the spectral peaks and information loss. Here we overcome the numerical scaling issues of the KBEs by developing a new scheme that picks up features beyond the mean-field level, demonstrates excellent agreement with exact results and has effectively linear scaling in the number of time-steps.

In this letter we present a route to include dynamical correlations for the time off-diagonal
Green’s function without the need to explicitly perform collision integrals. To compute the non-equilibrium spectral function one needs the off-diagonal components of the GF within a small neighborhood of the time-diagonal, centered around some time $t_p$, as shown schematically in Fig. \ref{fig:RT-DE_schematic}. Here, the scaling for computing the non-equilibrium spectral function is at best $O(\tilde N_t^2)$ , where  $\Tilde{N}_t$ is the number of time-points centered around the probe time $t_p$. In practice, $\Tilde{N}_t$ is dependent on the windowing function, $\mathcal{S}(t)$ which is typically Gaussian and further discussed in the supplemental information (SI)\cite{supp}. The time-resolved spectral function is expressed as,\cite{Freericks_2009}
\begin{equation}\label{eq:spectral_function}
        \mathcal{A}(\omega,t_p) = \int dt dt'\medspace \mathrm{e}^{-i\omega(t-t')}\mathcal{S}(t-t_p)\mathcal{S}(t'-t_p) \textrm{Tr}[G(t,t')].\\
\end{equation}

Where $\mathcal{S}(t)$ determines the energy and temporal resolution of $\mathcal{A}(\omega,t_p)$ \cite{Randi_2017}.  In practice $\Tilde{N}_t \ll N_t$ and is constant and determined solely by $\mathcal{S}(t)$ and the overall computational scaling is thus determined solely by the time-diagonal propagation.
\newline
\paragraph*{The real-time Dyson expansion (RT-DE)}
In the SI we show that\cite{supp}.
\begin{equation}\label{eq:dyson_eq}
   \begin{split}
        &G(12) = G^{\mathrm{MF}}(12) + \int\int d3 d4  \medspace G^{\mathrm{MF}}(13)\Sigma[ G^{\mathrm{MF}}(34)] G(42) \\&\equiv\left[i\partial_{t_1}- h^{\mathrm{MF}}[G^\mathrm{MF}(1)]\right]G(12)\\&\hspace{25mm} - \int d3 \medspace \Sigma[G^{\mathrm{MF}}(13)] G(32) = \delta(12) 
   \end{split}
\end{equation}

The first equation, gives the Dyson equation describing a one-shot correction on-top of a reference (arbitrary mean-field Hamiltonian (MFH)) for an arbitrary self-energy, $\Sigma(t,t')$\cite{Vacondio_2022}.  The second is exactly the equation we will solve in this paper using a real-time GF formalism. At equilibrium, our approach reduces to the standard Many-body perturbation theory (MBPT) correction on top of a MFH.  This typically leads to great improvement and even yields spectra containing signatures of strong dynamical correlations\cite{Mejuto_2022,Mejuto_2021,Vlcek_2019}.   Equation \eqref{eq:dyson_eq} builds an intuition behind the procedure: we use a mean-field method as an initial guess and we add a correction term that includes dynamical correlation effects through $\Sigma(t,t')$.  RT-DE thus represents a natural extension of the one-shot correction in MBPT to non-equilibrium settings.  

Using this form of the integro-differential equation for $G^{\mathrm{R}}(t,t')$ and the equations of motion for $G^{\mathrm{MF}}(t,t')$ we derive an ordinary differential equation (ODE) scheme that involves propagating $G^{\mathrm{R}}(t,t')$ along with an auxiliary two-body propagator.  A similar approach has also been applied to the time-diagonal GF in a linear scaling HF-GKBA scheme\cite{Schlunzen_2020,Joost_2020}.  Propagating the ODE scheme on the time off-diagonal for some $\Tilde{N_t}$ steps scales as $O(\Tilde{N}_t)$. Thus reconstructing the off-diagonal from $\Tilde{N}_t$ steps along the diagonal will indeed scale as $O(\Tilde{N}^2_t)$.  Furthermore, the time evolution is independent for each starting point, making this scheme extremely easy to parallelize in time coordinates. 

In the following, we only show the equations for $G^{\mathrm{<}}(t,t')$ however we perform the same procedure for $G^\mathrm{R}(t,t')$.  The details are in the SI, which includes the derivation of the RT-DE scheme\cite{supp}.

The  RT-DE involves solving the following equation for $G^{<}(t,t')$.
\begin{equation}
\begin{split}
\frac{\mathrm{d}G^{\mathrm{<}}(t,t')}{\mathrm{d}t} &= -i\left[h^{\mathrm{MF}}(t) G^{\mathrm{<}}(t,t') + I^{\mathrm{<}}(t,t')\right],\\
I^{\mathrm{<}}_{im}(t,t') &= -\sum_{klp}w_{iklp}(t) \mathcal{F}_{lpmk}(t,t').\\
\end{split}
\end{equation}
Here, the lesser collision integral, $I^{<}(t,t')$, depends on the interaction tensor and a two-body propagator $\mathcal{F}_{ijkl}$. By employing the approximation in equation \eqref{eq:dyson_eq} we introduce an ODE for $\mathcal{F}_{ijkl}$ given by

\begin{widetext}
\begin{equation}\label{RT-DE}
       \begin{split}
\frac{\mathrm{d}\mathcal{F}_{lpmk}(t,t')}{\mathrm{d}t} &=  \sum_{qrsj} w_{qrsj}(t)\Big{[} G^{>,\textrm{MF}}_{lq}(t)G^{<,\textrm{MF}}_{pr}(t)G^{>,\textrm{MF}}_{sk}(t) - G^{<,\textrm{MF}}_{lq}(t)G^{<,\textrm{MF}}_{pr}(t)G^{<,\textrm{MF}}_{sk}(t)\Big{]} G^{\mathrm{<}}_{jm}(t,t') \\&\hspace{50mm}-i\sum_{x}\Big{[} h^\mathrm{MF}_{lx}(t) \mathcal{F}_{xpmk}(t,t') + h^\mathrm{MF}_{px}(t) \mathcal{F}_{lxmk} (t,t') - \mathcal{F}_{lpmx}(t,t') h^\mathrm{MF}_{xk}(t)\Big{]}.\\
   \end{split}
\end{equation}
\end{widetext}

To fully demonstrate the the quality of this ``one-shot correction'' at equilibrium, we provide its numerical analysis in the SI, where we see excellent results\cite{supp}.  Additionally we discuss the failings of mean-field approaches like HF and the HF-GKBA

In the non-equilibrium setting the RT-DE requires a time evolution of $G^{<}(t,t)$ using a time-dependent MFH.    While our scheme is applicable to any particular choice of the reference static Hamiltonian, such as from a time-dependent density functional theory calculation\cite{ullrich_2012} and even those beyond mean-field e.g from some form of a downfolding procedure, we apply TD-HF on the time-diagonals for simplicity. This is discussed further in the SI\cite{supp}.  For completeness, we also show in \cite{supp} that the difference between exact diagonalization and TD-HF is relatively small on the diagonal.  The time-diagonal components are inserted in equation \eqref{RT-DE}. Starting from some $t'=t$ we can time-step in the $t$ variable in the range $t'\leq t\leq T_{\mathrm{max}}$.  Here we again emphasize an important advantage of the RT-DE scheme.  Since we can start from any point on the diagonal we choose to only compute those $\Tilde{N}_t^2$ off-diagonal terms within the probe window.

Finally, we we need to select an initial condition for $I^{<}(t,t)$ and $G^{<}(t,t)$ which, in principle, can break the equivalence to the standard perturbative expansion. This matter is discussed in more detail in the SI, as well as the initial conditions for the retarded component\cite{supp}.  For the results shown here we have taken $I^{<}(t,t) = 0$ and $G^{<}(t,t) = G^{<,\mathrm{MF}}(t)$. 
\newline
\paragraph*{Application to non-equilibrium problems}
We will now proceed by studying $\mathcal{A}(\omega,t_p)$ with our proposed scheme.  In this manuscript we only discuss comparisons of $\mathcal{A}(\omega,t_p)$ since this is the experimentally relevant quantity extracted from the time off-diagonal GF. Further, this serves to compress the information stored in the GF making our comparisons clearer.
\begin{equation}\label{eq:MB_ham}
\begin{split}
        \mathcal{H}  &= -J\sum_{\langle i,j\rangle}c^\dagger_ic_{j} + U\sum_{i}n_{i\uparrow}n_{i\downarrow} + \sum_{ij}h_{ij}^{\mathrm{N.E}}(t)c_{i}^\dagger c_j\\
\end{split}
\end{equation}

First, to demonstrate the accuracy of our method, we apply the RT-DE to non-equilibrium system and compare to TD-HF and exact diagonalization. We excite the system given by equation \eqref{eq:MB_ham} with a short wavelength pulse given by $h_{ij}^{\mathrm{N.E}}(t)=\delta_{ij}E\cos\left(\frac{\pi r_i}{2}\right)\exp\left({-\frac{(t-t_0)^2}{2T_p^2}}\right)$\cite{Reeves_2023}, where $r_i = \frac{N_s-1}{2} -i$ is the lattice vector for site $i$ and $N_s$ is the number of lattice sites.  Results for the lesser and retarded spectrum for dynamics induced by a long wavelength electric field and by a quench of the system are discussed in the SI\cite{supp}.
\begin{figure}
    \centering
\includegraphics[width=.4\textwidth]{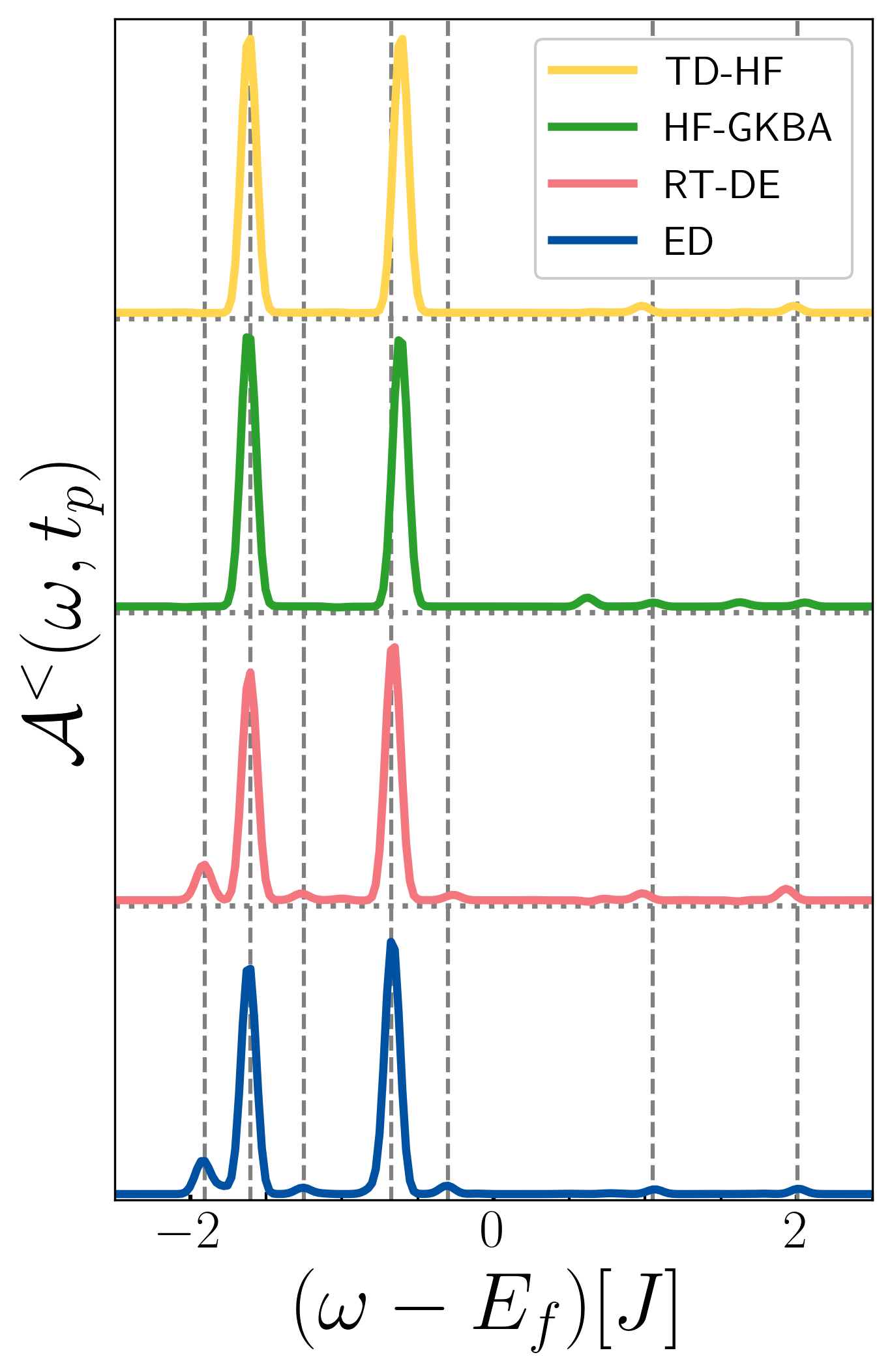}
    \caption{Time-resolved emission spectrum exact diagonalization (ED), RT-DE which is in excellent agreement with ED, and HF-GKBA that  mostly copies the TD Hartree Fock (TD-HF) spectrum. Calculated in the model given in equation \eqref{eq:MB_ham} with the following parameters: $U=1.0J$,  $N_s = 4$, $E = 0.5J$, $t_0 = 5J^{-1}$ and $T_p = 0.5J^{-1}$ after time evolving to $T_{\mathrm{max}} = 200J^{-1}$.  The probe width is taken to be $\delta = 10$ and $t_p = 100J^{-1}$.}
    \label{fig:excited_state}
\end{figure}

Examining Fig. \ref{fig:excited_state} we see the RT-DE is in excellent agreement with the exact result for the lesser component of the $\mathcal{A}(\omega,t_p)$. It picks up features around $\omega \approx -1.2J$ and $\omega \approx -0.4J$ that are completely missed by TD-HF and HF-GKBA, which mostly copies the TD-HF features.  Notably, neither of these peaks is present in equilibrium\cite{supp}. This shows the RT-DE can pick up emergent features that arise purely due to non-equilibrium effects in the system.  For the peaks above the Fermi energy,the RT-DE, HF-GKBA and TD-HF pick up the peaks at $\omega \approx 1.0J$ and $2.0J$. TD-HF and the RT-DE methods slightly underestimate the energy of the peak around $\omega \approx 1.0J$ by the same amount. Interestingly, the peak near $\omega \approx 2.0J$ is capture slightly better with TD-HF than with the RT-DE. This is related to the initial condition for the lesser component. Given that the initial condition for $G^{<}(t,t')$ can be fixed in different ways. Further optimization and constraining the particular choices will be explored elsewhere. However, this boundary condition affects only the details of the electronic distribution in dynamically renormalized states. The band renormalization and emergence of new spectral peaks is however captured perfectly using the knowledge of the $G^R$ alone, which is free of any uncertainty. Finally, we note the appearance of two additional peaks in the HF-GKBA result at $\omega\approx 0.5J$ and $\omega\approx 1.6J$

For completeness we provide discussion of the full spectrum from $G^{\mathrm{R}}(t,t')$ in the SI\cite{supp}.

In our second example we explore a realistic model of an insulator out of equilibrium. We study the dynamics of a long-range interacting two band Hubbard model with periodic boundary conditions at half filling under a strong optical excitation.  The full Hamiltonian is given by
\begin{equation}\label{eq:multiband_ham}
\begin{split}
        \mathcal{H} &= \sum_{\alpha,\beta\in \{c,v\}}\sum_{ \langle i,j\rangle,\sigma} h^{\alpha\beta}_{ij}(t)c^{\dagger\alpha}_{i,\sigma}c^{\beta}_{j,\sigma} + U\sum_{i,\alpha}n^\alpha_{i\uparrow}n^{\alpha}_{i\downarrow} \\&\hspace{25mm}+U\sum_{i}n_{ic}n_{iv} + \gamma U\sum_{i < j\atop\alpha\leq\beta} \frac{n_{i\alpha} n_{j\beta}}{|\Vec{r}_i - \Vec{r}_j|}.\\
        h^{\alpha\beta}_{ij}(t) &= J\delta_{\alpha\beta}\delta_{\langle i,j\rangle}(\delta_{\alpha v }-\delta_{\alpha c})+ \epsilon_\alpha\delta_{\alpha\beta} \\&\hspace{15mm} + \delta_{ij}(1-\delta_{\alpha\beta})
        E\cos(\omega_p(t-t_0)) \mathrm{e}^{-\frac{(t-t_0)^2}{2T_p^2}}
\end{split}
\end{equation}
\begin{figure*}
    \centering    
\includegraphics[width=\textwidth]{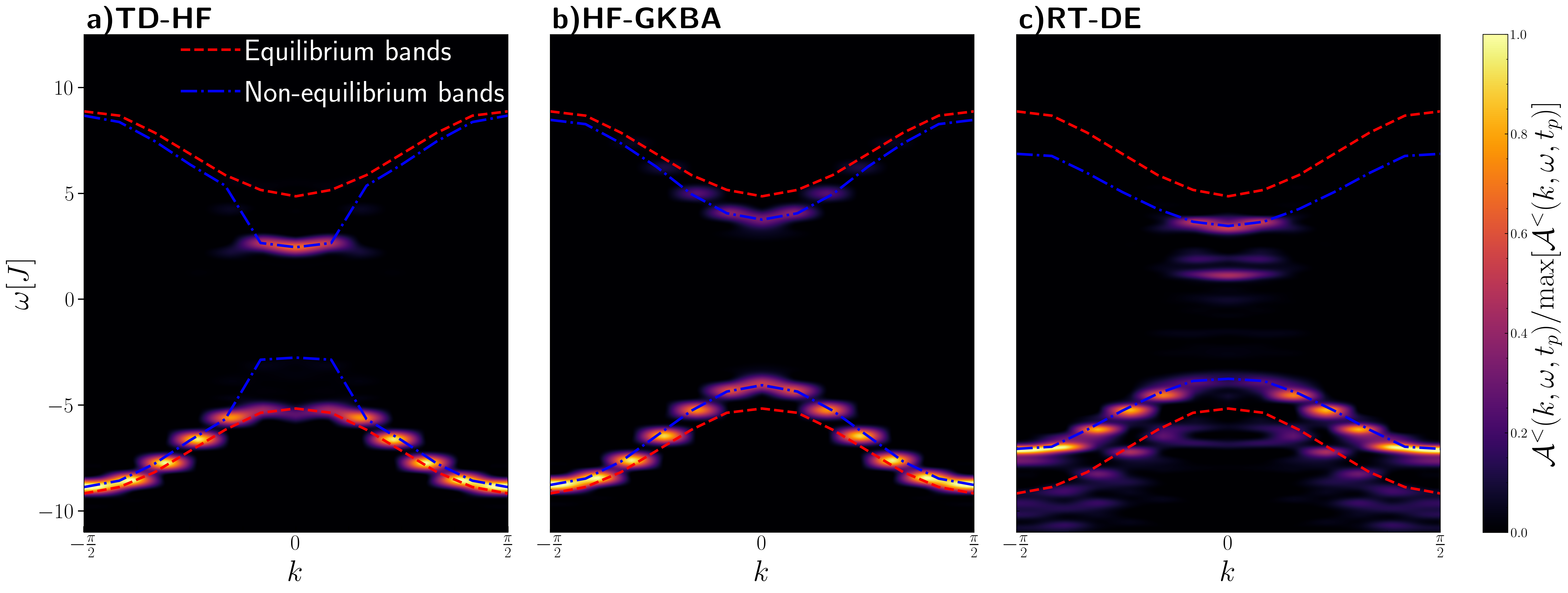}
    \caption{Non-equilibrium time-resolved emission spectrum for the model given in equation \eqref{eq:multiband_ham} with 12 sites evolved until time $T_{\mathrm{max}} = 100J^{-1}$. Panels a), b) and c) show the spectra computed with TD-HF, the HF-GKBA and the RT-DE respectively.  The model parameters are $U=4.0J$, $\gamma=0.5$, $\epsilon_v=-30J$,  $\epsilon_c=-20J$, $E=2.5J$, $\omega_p=7.5J$, $t_0=10J^{-1}$, $T_p=1.8J^{-1}$.  The probe width used in \eqref{eq:spectral_function} is $\delta = 5$ and the probe is centered at $t_p = 90J^{-1}$.  A prominent excitonic feature is seen for the RT-DE results just below the renormalized conduction band; further, a visible band satellite is visible at the bottom of the bandstructure. }
    \label{fig:periodic_system}
\end{figure*}
The results are shown in Fig. \ref{fig:periodic_system}.  We compare TD-HF, the HF-GKBA and the RT-DE using the following model parameters: $U=4.0J$, $\epsilon_v=-30J$,  $\epsilon_c=-20J$, $\gamma = 0.5,E=2.5J$, $\omega_p=7.5J$,  $t_0=10J^{-1}$ and $T_p=1.8J^{-1}$.  The excitation frequency (pump), $\omega_p$ is below the equilibrium gap ($\sim 10.0J$) but it is chosen such that excitons can form in the system.  We also remark that at equilibrium the three methods produce practically identical band structures, shown as red dashed lines in Fig. \ref{fig:periodic_system}.  This allows us to see how each method performs, purely from changes due to the system leaving equilibrium.

The periodic system is too large for benchmarking against exact results; hence, this example serves primarily as a demonstration of the ability of the RT-DE to be applied to relatively large periodic systems. Even in the absence
of the exact reference, we compare the qualitative features of $\mathcal{A}(\omega,t_p)$ and determine the physical origins of the
dynamical features that are produced by RT-DE (but not by the other
methods).

Fig. \ref{fig:periodic_system} a) shows $\mathcal{A}(\omega,t_p)$ for TD-HF.  The full non-equilibrium band structure is shown in blue dash-dotted lines.  Due to the lack of dynamical self-energy (yielding possibly multiple quasiparticle solutions) no new (satellite) states can appear in the TD-HF result.  The figure shows electrons are excited across the gap and into the conduction band around $k=0$.   This change in the  electron distribution leads to a renormalization of the band gap, which is governed by the changes in the orbital occupancies and related to changes in the time dependent density matrix. In practice, occupied states are shifted down in energy creating a relatively sharp discontinuity in the upper band which becomes partially filled around $k=0$.  The concentration of electrons (holes) at the the conduction (valence) band extrema and the mean-field nature of TD-HF means the renormalization mainly effects the $k=0$ point, leading to a significant stretching of the bands.  As expected the lack of two-body correlations in TD-HF means we observe no novel features in $\mathcal{A}(\omega,t_p)$ which could be interpreted as emerging excitonic states beyond the band stretching.

In Fig. \ref{fig:periodic_system} b) we again show $\mathcal{A}(\omega,t_p)$, now computed using the HF-GKBA.  This result shares features common to TD-HF as we expect because it is dominated by the off-diagonal terms dominated by HF.  Firstly, while the HF signatures are diminished, we see a similar stretching of the bands relative to their equilibrium value. One reason for this change comes from an inconsistency between the treatment of the time-diagonal and off-diagonal terms in HF-GKBA, as the HF-GKBA density matrix, $\rho^{\mathrm{HF-GKBA}}(t)$, enters the HF equation for the off-diagonal instead of $\rho^{\mathrm{HF}}(t)$. The density matrix is thus projected on to a set of excited states of the HF Hamiltonian.     Such an inconsistency makes it a priori difficult to interpret HF-GKBA spectral results.  Importantly the HF-GKBA again lacks additional excitonic features. The picture is thus consistent with TD-HF (albeit with distinct band renormalization ).  

Finally we turn to Fig. \ref{fig:periodic_system} c), showing the RT-DE results. The results differ markedly and show emergent features. For the main renormalized bands we see a renormalization of the band positions compared to equilibrium.  This affects both the $k\neq0$ and $k=0$ points very similarly resulting in almost a rigid shift of the bands, compared to the stretching seen in Fig \ref{fig:periodic_system} a) and b). This is consistent with renormalized (increased) screening in the photoexcited system affecting the entire band structure and not only the small number states that become occupied by photo-excitation. 

In addition, there are two more striking differences.  The first is  a peak appearing just below the renormalized conduction band, consistent with an in-gap electronic state, i.e., an exciton.  Secondly we see the appearance of a satellite band appearing just below the position of the equilibrium valence band.  Similarly to the finite system in Fig.~\ref{fig:excited_state} the inclusion of dynamical correlations through $I^<(t,t')$ allows for the formation of additional spectral features. This is consistent with satellites emerging due to coupling to a bosonic excitation in the system. Here, the satellite is displaced by energy equivalent to the distance between the excitonic feature and the minimum of the renormalized conduction state. Potentially, one can interpret this as an ``excitonic replica", i.e., simulaneous QP formation and exciton dissociation, similar to other bosonic replicas\cite{Mejuto_2021}. The exact interpretation of such features are left for future research, but we note that these are consistent with the exact solutions made for finite systems studied above.  
\newline
\paragraph*{Conclusions and outlook}
For our tests in finite systems the results for the RT-DE in and out of equilibrium are in excellent agreement with exact results. Not only does the RT-DE correct features picked up by TD-HF and the HF-GKBA but it also captures satellite features that both other methods miss entirely.  It is important to highlight that these excellent results are produced using the second Born approximation (representing one of the simplest self-energy forms). Thus, the results and comparisons to exact diagonalization given in this manuscript serve as compelling a proof of principle of the RT-DE.  Already our method can capture dynamical correlation effects such as plasmons to a high degree of accuracy and it is natural to expect that extending to more advanced self-energy approximations will only further improve upon the results presented here, an area that is currently being pursued.  To  show the generality of our approach we have provided a derivation of the RT-DE for the $GW$ self-energy in the SI[26], it's implementation, testing and comparison is left to a later publication.

For our periodic system, the RT-DE provides physically meaningful results.  Importantly by including two-body correlations through $\Sigma(t,t')$ we are able to observe an emergent excitonic peak appearing in $\mathcal{A}(\omega,t_p)$ and energetically well separated from the quasiparticle band. 
 
In conclusion the RT-DE is an efficient scheme for studying the non-equilibrium spectral properties of quantum systems,  offering accuracy and a huge reduction in cost relative to the full KBEs.  While KBEs and RT-DE will scale similarly in system size, by reducing the cost from cubic to linear in the number of time-steps, we overcome one of the major limitations of the KBEs that has prevented them being widely applied to non-equilibrium first principles simulations.  Application of this scheme to a more realistic system and comparison of the results to experimental measurements as a validation is underway. 
Finally, we note that in principle the RT-DE can be combined with existing real-time MBPT methods and their extension to more advanced self-energy approximations, such as $GW\Gamma$\cite{Mejuto_2022}.  This extension is currently underway and will open the doors to the accurate and efficient simulation of time-resolved photo-emission experiments.
\newline
\paragraph*{Acknowledgements}
This material is based upon work supported by the U.S. Department of Energy, Office of Science, Office of Advanced Scientific Computing Research and Office of Basic Energy Sciences, Scientific Discovery through Advanced Computing (SciDAC) program under Award Number DE-SC0022198.  This research used resources of the National Energy Research Scientific Computing Center, a DOE Office of Science User Facility supported by the Office of Science of the U.S. Department of Energy under Contract No. DE-AC02-05CH11231 using NERSC award BES-ERCAP0029462

\bibliography{Bib}

\begin{thebibliography}{44}%
\makeatletter
\providecommand \@ifxundefined [1]{%
 \@ifx{#1\undefined}
}%
\providecommand \@ifnum [1]{%
 \ifnum #1\expandafter \@firstoftwo
 \else \expandafter \@secondoftwo
 \fi
}%
\providecommand \@ifx [1]{%
 \ifx #1\expandafter \@firstoftwo
 \else \expandafter \@secondoftwo
 \fi
}%
\providecommand \natexlab [1]{#1}%
\providecommand \enquote  [1]{``#1''}%
\providecommand \bibnamefont  [1]{#1}%
\providecommand \bibfnamefont [1]{#1}%
\providecommand \citenamefont [1]{#1}%
\providecommand \href@noop [0]{\@secondoftwo}%
\providecommand \href [0]{\begingroup \@sanitize@url \@href}%
\providecommand \@href[1]{\@@startlink{#1}\@@href}%
\providecommand \@@href[1]{\endgroup#1\@@endlink}%
\providecommand \@sanitize@url [0]{\catcode `\\12\catcode `\$12\catcode `\&12\catcode `\#12\catcode `\^12\catcode `\_12\catcode `\%12\relax}%
\providecommand \@@startlink[1]{}%
\providecommand \@@endlink[0]{}%
\providecommand \url  [0]{\begingroup\@sanitize@url \@url }%
\providecommand \@url [1]{\endgroup\@href {#1}{\urlprefix }}%
\providecommand \urlprefix  [0]{URL }%
\providecommand \Eprint [0]{\href }%
\providecommand \doibase [0]{https://doi.org/}%
\providecommand \selectlanguage [0]{\@gobble}%
\providecommand \bibinfo  [0]{\@secondoftwo}%
\providecommand \bibfield  [0]{\@secondoftwo}%
\providecommand \translation [1]{[#1]}%
\providecommand \BibitemOpen [0]{}%
\providecommand \bibitemStop [0]{}%
\providecommand \bibitemNoStop [0]{.\EOS\space}%
\providecommand \EOS [0]{\spacefactor3000\relax}%
\providecommand \BibitemShut  [1]{\csname bibitem#1\endcsname}%
\let\auto@bib@innerbib\@empty
\bibitem [{\citenamefont {Beebe}\ \emph {et~al.}(2017)\citenamefont {Beebe}, \citenamefont {Klopf}, \citenamefont {Wang}, \citenamefont {Kittiwatanakul}, \citenamefont {Lu}, \citenamefont {Wolf},\ and\ \citenamefont {Lukaszew}}]{Beebe_2017}%
  \BibitemOpen
  \bibfield  {author} {\bibinfo {author} {\bibfnamefont {M.~R.}\ \bibnamefont {Beebe}}, \bibinfo {author} {\bibfnamefont {J.~M.}\ \bibnamefont {Klopf}}, \bibinfo {author} {\bibfnamefont {Y.}~\bibnamefont {Wang}}, \bibinfo {author} {\bibfnamefont {S.}~\bibnamefont {Kittiwatanakul}}, \bibinfo {author} {\bibfnamefont {J.}~\bibnamefont {Lu}}, \bibinfo {author} {\bibfnamefont {S.~A.}\ \bibnamefont {Wolf}},\ and\ \bibinfo {author} {\bibfnamefont {R.~A.}\ \bibnamefont {Lukaszew}},\ }\bibfield  {title} {\bibinfo {title} {{Time-resolved light-induced insulator-metal transition in niobium dioxide and vanadium dioxide thin films}},\ }\href {https://doi.org/10.1364/OME.7.000213} {\bibfield  {journal} {\bibinfo  {journal} {Opt. Mater. Express}\ }\textbf {\bibinfo {volume} {7}},\ \bibinfo {pages} {213} (\bibinfo {year} {2017})}\BibitemShut {NoStop}%
\bibitem [{\citenamefont {Disa}\ \emph {et~al.}(2023)\citenamefont {Disa}, \citenamefont {Curtis}, \citenamefont {Fechner}, \citenamefont {Liu}, \citenamefont {von Hoegen}, \citenamefont {F{\"o}rst}, \citenamefont {Nova}, \citenamefont {Narang}, \citenamefont {Maljuk}, \citenamefont {Boris}, \citenamefont {Keimer},\ and\ \citenamefont {Cavalleri}}]{Disa_2023}%
  \BibitemOpen
  \bibfield  {author} {\bibinfo {author} {\bibfnamefont {A.~S.}\ \bibnamefont {Disa}}, \bibinfo {author} {\bibfnamefont {J.}~\bibnamefont {Curtis}}, \bibinfo {author} {\bibfnamefont {M.}~\bibnamefont {Fechner}}, \bibinfo {author} {\bibfnamefont {A.}~\bibnamefont {Liu}}, \bibinfo {author} {\bibfnamefont {A.}~\bibnamefont {von Hoegen}}, \bibinfo {author} {\bibfnamefont {M.}~\bibnamefont {F{\"o}rst}}, \bibinfo {author} {\bibfnamefont {T.~F.}\ \bibnamefont {Nova}}, \bibinfo {author} {\bibfnamefont {P.}~\bibnamefont {Narang}}, \bibinfo {author} {\bibfnamefont {A.}~\bibnamefont {Maljuk}}, \bibinfo {author} {\bibfnamefont {A.~V.}\ \bibnamefont {Boris}}, \bibinfo {author} {\bibfnamefont {B.}~\bibnamefont {Keimer}},\ and\ \bibinfo {author} {\bibfnamefont {A.}~\bibnamefont {Cavalleri}},\ }\bibfield  {title} {\bibinfo {title} {{Photo-induced high-temperature ferromagnetism in YTiO3}},\ }\href {https://doi.org/10.1038/s41586-023-05853-8} {\bibfield  {journal} {\bibinfo  {journal} {Nature}\ }\textbf {\bibinfo {volume}
  {617}},\ \bibinfo {pages} {73} (\bibinfo {year} {2023})}\BibitemShut {NoStop}%
\bibitem [{\citenamefont {Dong}\ \emph {et~al.}(2021)\citenamefont {Dong}, \citenamefont {Puppin}, \citenamefont {Pincelli}, \citenamefont {Beaulieu}, \citenamefont {Christiansen}, \citenamefont {Hübener}, \citenamefont {Nicholson}, \citenamefont {Xian}, \citenamefont {Dendzik}, \citenamefont {Deng}, \citenamefont {Windsor}, \citenamefont {Selig}, \citenamefont {Malic}, \citenamefont {Rubio}, \citenamefont {Knorr}, \citenamefont {Wolf}, \citenamefont {Rettig},\ and\ \citenamefont {Ernstorfer}}]{Dong_2021}%
  \BibitemOpen
  \bibfield  {author} {\bibinfo {author} {\bibfnamefont {S.}~\bibnamefont {Dong}}, \bibinfo {author} {\bibfnamefont {M.}~\bibnamefont {Puppin}}, \bibinfo {author} {\bibfnamefont {T.}~\bibnamefont {Pincelli}}, \bibinfo {author} {\bibfnamefont {S.}~\bibnamefont {Beaulieu}}, \bibinfo {author} {\bibfnamefont {D.}~\bibnamefont {Christiansen}}, \bibinfo {author} {\bibfnamefont {H.}~\bibnamefont {Hübener}}, \bibinfo {author} {\bibfnamefont {C.~W.}\ \bibnamefont {Nicholson}}, \bibinfo {author} {\bibfnamefont {R.~P.}\ \bibnamefont {Xian}}, \bibinfo {author} {\bibfnamefont {M.}~\bibnamefont {Dendzik}}, \bibinfo {author} {\bibfnamefont {Y.}~\bibnamefont {Deng}}, \bibinfo {author} {\bibfnamefont {Y.~W.}\ \bibnamefont {Windsor}}, \bibinfo {author} {\bibfnamefont {M.}~\bibnamefont {Selig}}, \bibinfo {author} {\bibfnamefont {E.}~\bibnamefont {Malic}}, \bibinfo {author} {\bibfnamefont {A.}~\bibnamefont {Rubio}}, \bibinfo {author} {\bibfnamefont {A.}~\bibnamefont {Knorr}}, \bibinfo {author} {\bibfnamefont {M.}~\bibnamefont
  {Wolf}}, \bibinfo {author} {\bibfnamefont {L.}~\bibnamefont {Rettig}},\ and\ \bibinfo {author} {\bibfnamefont {R.}~\bibnamefont {Ernstorfer}},\ }\bibfield  {title} {\bibinfo {title} {{Direct measurement of key exciton properties: Energy, dynamics, and spatial distribution of the wave function}},\ }\href {https://doi.org/https://doi.org/10.1002/ntls.10010} {\bibfield  {journal} {\bibinfo  {journal} {Natural Sciences}\ }\textbf {\bibinfo {volume} {1}},\ \bibinfo {pages} {e10010} (\bibinfo {year} {2021})}\BibitemShut {NoStop}%
\bibitem [{\citenamefont {Bao}\ \emph {et~al.}(2022)\citenamefont {Bao}, \citenamefont {Tang}, \citenamefont {Sun},\ and\ \citenamefont {Zhou}}]{Bao_2022}%
  \BibitemOpen
  \bibfield  {author} {\bibinfo {author} {\bibfnamefont {C.}~\bibnamefont {Bao}}, \bibinfo {author} {\bibfnamefont {P.}~\bibnamefont {Tang}}, \bibinfo {author} {\bibfnamefont {D.}~\bibnamefont {Sun}},\ and\ \bibinfo {author} {\bibfnamefont {S.}~\bibnamefont {Zhou}},\ }\bibfield  {title} {\bibinfo {title} {{Light-induced emergent phenomena in 2D materials and topological materials}},\ }\href {https://doi.org/10.1038/s42254-021-00388-1} {\bibfield  {journal} {\bibinfo  {journal} {Nature Reviews Physics}\ }\textbf {\bibinfo {volume} {4}},\ \bibinfo {pages} {33} (\bibinfo {year} {2022})}\BibitemShut {NoStop}%
\bibitem [{\citenamefont {Nuske}\ \emph {et~al.}(2020)\citenamefont {Nuske}, \citenamefont {Broers}, \citenamefont {Schulte}, \citenamefont {Jotzu}, \citenamefont {Sato}, \citenamefont {Cavalleri}, \citenamefont {Rubio}, \citenamefont {McIver},\ and\ \citenamefont {Mathey}}]{Nuske_2020}%
  \BibitemOpen
  \bibfield  {author} {\bibinfo {author} {\bibfnamefont {M.}~\bibnamefont {Nuske}}, \bibinfo {author} {\bibfnamefont {L.}~\bibnamefont {Broers}}, \bibinfo {author} {\bibfnamefont {B.}~\bibnamefont {Schulte}}, \bibinfo {author} {\bibfnamefont {G.}~\bibnamefont {Jotzu}}, \bibinfo {author} {\bibfnamefont {S.~A.}\ \bibnamefont {Sato}}, \bibinfo {author} {\bibfnamefont {A.}~\bibnamefont {Cavalleri}}, \bibinfo {author} {\bibfnamefont {A.}~\bibnamefont {Rubio}}, \bibinfo {author} {\bibfnamefont {J.~W.}\ \bibnamefont {McIver}},\ and\ \bibinfo {author} {\bibfnamefont {L.}~\bibnamefont {Mathey}},\ }\bibfield  {title} {\bibinfo {title} {{Floquet dynamics in light-driven solids}},\ }\href {https://doi.org/10.1103/PhysRevResearch.2.043408} {\bibfield  {journal} {\bibinfo  {journal} {Phys. Rev. Res.}\ }\textbf {\bibinfo {volume} {2}},\ \bibinfo {pages} {043408} (\bibinfo {year} {2020})}\BibitemShut {NoStop}%
\bibitem [{\citenamefont {Zhou}\ \emph {et~al.}(2023)\citenamefont {Zhou}, \citenamefont {Bao}, \citenamefont {Fan}, \citenamefont {Wang}, \citenamefont {Zhong}, \citenamefont {Zhang}, \citenamefont {Tang}, \citenamefont {Duan},\ and\ \citenamefont {Zhou}}]{Zhou_2023}%
  \BibitemOpen
  \bibfield  {author} {\bibinfo {author} {\bibfnamefont {S.}~\bibnamefont {Zhou}}, \bibinfo {author} {\bibfnamefont {C.}~\bibnamefont {Bao}}, \bibinfo {author} {\bibfnamefont {B.}~\bibnamefont {Fan}}, \bibinfo {author} {\bibfnamefont {F.}~\bibnamefont {Wang}}, \bibinfo {author} {\bibfnamefont {H.}~\bibnamefont {Zhong}}, \bibinfo {author} {\bibfnamefont {H.}~\bibnamefont {Zhang}}, \bibinfo {author} {\bibfnamefont {P.}~\bibnamefont {Tang}}, \bibinfo {author} {\bibfnamefont {W.}~\bibnamefont {Duan}},\ and\ \bibinfo {author} {\bibfnamefont {S.}~\bibnamefont {Zhou}},\ }\bibfield  {title} {\bibinfo {title} {Floquet engineering of black phosphorus upon below-gap pumping},\ }\href {https://doi.org/10.1103/PhysRevLett.131.116401} {\bibfield  {journal} {\bibinfo  {journal} {Phys. Rev. Lett.}\ }\textbf {\bibinfo {volume} {131}},\ \bibinfo {pages} {116401} (\bibinfo {year} {2023})}\BibitemShut {NoStop}%
\bibitem [{\citenamefont {Sie}\ \emph {et~al.}(2019)\citenamefont {Sie}, \citenamefont {Rohwer}, \citenamefont {Lee},\ and\ \citenamefont {Gedik}}]{Sie_2019}%
  \BibitemOpen
  \bibfield  {author} {\bibinfo {author} {\bibfnamefont {E.~J.}\ \bibnamefont {Sie}}, \bibinfo {author} {\bibfnamefont {T.}~\bibnamefont {Rohwer}}, \bibinfo {author} {\bibfnamefont {C.}~\bibnamefont {Lee}},\ and\ \bibinfo {author} {\bibfnamefont {N.}~\bibnamefont {Gedik}},\ }\bibfield  {title} {\bibinfo {title} {{Time-resolved XUV ARPES with tunable 24--33{\thinspace}eV laser pulses at 30{\thinspace}meV resolution}},\ }\href {https://doi.org/10.1038/s41467-019-11492-3} {\bibfield  {journal} {\bibinfo  {journal} {Nature Communications}\ }\textbf {\bibinfo {volume} {10}},\ \bibinfo {pages} {3535} (\bibinfo {year} {2019})}\BibitemShut {NoStop}%
\bibitem [{\citenamefont {Karni}\ \emph {et~al.}(2023)\citenamefont {Karni}, \citenamefont {Esin},\ and\ \citenamefont {Dani}}]{Karni_2023}%
  \BibitemOpen
  \bibfield  {author} {\bibinfo {author} {\bibfnamefont {O.}~\bibnamefont {Karni}}, \bibinfo {author} {\bibfnamefont {I.}~\bibnamefont {Esin}},\ and\ \bibinfo {author} {\bibfnamefont {K.~M.}\ \bibnamefont {Dani}},\ }\bibfield  {title} {\bibinfo {title} {{Through the Lens of a Momentum Microscope: Viewing Light-Induced Quantum Phenomena in 2D Materials}},\ }\href {https://doi.org/https://doi.org/10.1002/adma.202204120} {\bibfield  {journal} {\bibinfo  {journal} {Advanced Materials}\ }\textbf {\bibinfo {volume} {35}},\ \bibinfo {pages} {2204120} (\bibinfo {year} {2023})}\BibitemShut {NoStop}%
\bibitem [{\citenamefont {Man}\ \emph {et~al.}(2021)\citenamefont {Man}, \citenamefont {Madéo}, \citenamefont {Sahoo}, \citenamefont {Xie}, \citenamefont {Campbell}, \citenamefont {Pareek}, \citenamefont {Karmakar}, \citenamefont {Wong}, \citenamefont {Al-Mahboob}, \citenamefont {Chan}, \citenamefont {Bacon}, \citenamefont {Zhu}, \citenamefont {Abdelrasoul}, \citenamefont {Li}, \citenamefont {Heinz}, \citenamefont {da~Jornada}, \citenamefont {Cao},\ and\ \citenamefont {Dani}}]{Man_2021}%
  \BibitemOpen
  \bibfield  {author} {\bibinfo {author} {\bibfnamefont {M.~K.~L.}\ \bibnamefont {Man}}, \bibinfo {author} {\bibfnamefont {J.}~\bibnamefont {Madéo}}, \bibinfo {author} {\bibfnamefont {C.}~\bibnamefont {Sahoo}}, \bibinfo {author} {\bibfnamefont {K.}~\bibnamefont {Xie}}, \bibinfo {author} {\bibfnamefont {M.}~\bibnamefont {Campbell}}, \bibinfo {author} {\bibfnamefont {V.}~\bibnamefont {Pareek}}, \bibinfo {author} {\bibfnamefont {A.}~\bibnamefont {Karmakar}}, \bibinfo {author} {\bibfnamefont {E.~L.}\ \bibnamefont {Wong}}, \bibinfo {author} {\bibfnamefont {A.}~\bibnamefont {Al-Mahboob}}, \bibinfo {author} {\bibfnamefont {N.~S.}\ \bibnamefont {Chan}}, \bibinfo {author} {\bibfnamefont {D.~R.}\ \bibnamefont {Bacon}}, \bibinfo {author} {\bibfnamefont {X.}~\bibnamefont {Zhu}}, \bibinfo {author} {\bibfnamefont {M.~M.~M.}\ \bibnamefont {Abdelrasoul}}, \bibinfo {author} {\bibfnamefont {X.}~\bibnamefont {Li}}, \bibinfo {author} {\bibfnamefont {T.~F.}\ \bibnamefont {Heinz}}, \bibinfo {author} {\bibfnamefont {F.~H.}\
  \bibnamefont {da~Jornada}}, \bibinfo {author} {\bibfnamefont {T.}~\bibnamefont {Cao}},\ and\ \bibinfo {author} {\bibfnamefont {K.~M.}\ \bibnamefont {Dani}},\ }\bibfield  {title} {\bibinfo {title} {{Experimental measurement of the intrinsic excitonic wave function}},\ }\href {https://doi.org/10.1126/sciadv.abg0192} {\bibfield  {journal} {\bibinfo  {journal} {Science Advances}\ }\textbf {\bibinfo {volume} {7}},\ \bibinfo {pages} {eabg0192} (\bibinfo {year} {2021})}\BibitemShut {NoStop}%
\bibitem [{\citenamefont {Perfetto}\ \emph {et~al.}(2022)\citenamefont {Perfetto}, \citenamefont {Pavlyukh},\ and\ \citenamefont {Stefanucci}}]{Perfetto_2022}%
  \BibitemOpen
  \bibfield  {author} {\bibinfo {author} {\bibfnamefont {E.}~\bibnamefont {Perfetto}}, \bibinfo {author} {\bibfnamefont {Y.}~\bibnamefont {Pavlyukh}},\ and\ \bibinfo {author} {\bibfnamefont {G.}~\bibnamefont {Stefanucci}},\ }\bibfield  {title} {\bibinfo {title} {{Real-Time $GW$: Toward an Ab Initio Description of the Ultrafast Carrier and Exciton Dynamics in Two-Dimensional Materials}},\ }\href {https://doi.org/10.1103/PhysRevLett.128.016801} {\bibfield  {journal} {\bibinfo  {journal} {Phys. Rev. Lett.}\ }\textbf {\bibinfo {volume} {128}},\ \bibinfo {pages} {016801} (\bibinfo {year} {2022})}\BibitemShut {NoStop}%
\bibitem [{\citenamefont {Sun}\ \emph {et~al.}(2021)\citenamefont {Sun}, \citenamefont {Lee}, \citenamefont {Kononov}, \citenamefont {Schleife},\ and\ \citenamefont {Ullrich}}]{Sun_2021}%
  \BibitemOpen
  \bibfield  {author} {\bibinfo {author} {\bibfnamefont {J.}~\bibnamefont {Sun}}, \bibinfo {author} {\bibfnamefont {C.-W.}\ \bibnamefont {Lee}}, \bibinfo {author} {\bibfnamefont {A.}~\bibnamefont {Kononov}}, \bibinfo {author} {\bibfnamefont {A.}~\bibnamefont {Schleife}},\ and\ \bibinfo {author} {\bibfnamefont {C.~A.}\ \bibnamefont {Ullrich}},\ }\bibfield  {title} {\bibinfo {title} {{Real-Time Exciton Dynamics with Time-Dependent Density-Functional Theory}},\ }\href {https://doi.org/10.1103/PhysRevLett.127.077401} {\bibfield  {journal} {\bibinfo  {journal} {Phys. Rev. Lett.}\ }\textbf {\bibinfo {volume} {127}},\ \bibinfo {pages} {077401} (\bibinfo {year} {2021})}\BibitemShut {NoStop}%
\bibitem [{\citenamefont {Karlsson}\ \emph {et~al.}(2021)\citenamefont {Karlsson}, \citenamefont {van Leeuwen}, \citenamefont {Pavlyukh}, \citenamefont {Perfetto},\ and\ \citenamefont {Stefanucci}}]{Karlsson_2021}%
  \BibitemOpen
  \bibfield  {author} {\bibinfo {author} {\bibfnamefont {D.}~\bibnamefont {Karlsson}}, \bibinfo {author} {\bibfnamefont {R.}~\bibnamefont {van Leeuwen}}, \bibinfo {author} {\bibfnamefont {Y.}~\bibnamefont {Pavlyukh}}, \bibinfo {author} {\bibfnamefont {E.}~\bibnamefont {Perfetto}},\ and\ \bibinfo {author} {\bibfnamefont {G.}~\bibnamefont {Stefanucci}},\ }\bibfield  {title} {\bibinfo {title} {{Fast Green's Function Method for Ultrafast Electron-Boson Dynamics}},\ }\href {https://doi.org/10.1103/PhysRevLett.127.036402} {\bibfield  {journal} {\bibinfo  {journal} {Phys. Rev. Lett.}\ }\textbf {\bibinfo {volume} {127}},\ \bibinfo {pages} {036402} (\bibinfo {year} {2021})}\BibitemShut {NoStop}%
\bibitem [{\citenamefont {Schl\"unzen}\ \emph {et~al.}(2020)\citenamefont {Schl\"unzen}, \citenamefont {Joost},\ and\ \citenamefont {Bonitz}}]{Schlunzen_2020}%
  \BibitemOpen
  \bibfield  {author} {\bibinfo {author} {\bibfnamefont {N.}~\bibnamefont {Schl\"unzen}}, \bibinfo {author} {\bibfnamefont {J.-P.}\ \bibnamefont {Joost}},\ and\ \bibinfo {author} {\bibfnamefont {M.}~\bibnamefont {Bonitz}},\ }\bibfield  {title} {\bibinfo {title} {{Achieving the Scaling Limit for Nonequilibrium {G}reen Functions Simulations}},\ }\href {https://doi.org/10.1103/PhysRevLett.124.076601} {\bibfield  {journal} {\bibinfo  {journal} {Phys. Rev. Lett.}\ }\textbf {\bibinfo {volume} {124}},\ \bibinfo {pages} {076601} (\bibinfo {year} {2020})}\BibitemShut {NoStop}%
\bibitem [{\citenamefont {Kaye}\ and\ \citenamefont {Golež}(2021)}]{Kaye_2021}%
  \BibitemOpen
  \bibfield  {author} {\bibinfo {author} {\bibfnamefont {J.}~\bibnamefont {Kaye}}\ and\ \bibinfo {author} {\bibfnamefont {D.}~\bibnamefont {Golež}},\ }\bibfield  {title} {\bibinfo {title} {{Low rank compression in the numerical solution of the nonequilibrium Dyson equation}},\ }\href {https://doi.org/10.21468/SciPostPhys.10.4.091} {\bibfield  {journal} {\bibinfo  {journal} {SciPost Phys.}\ }\textbf {\bibinfo {volume} {10}},\ \bibinfo {pages} {091} (\bibinfo {year} {2021})}\BibitemShut {NoStop}%
\bibitem [{\citenamefont {Lian}\ \emph {et~al.}(2018)\citenamefont {Lian}, \citenamefont {Guan}, \citenamefont {Hu}, \citenamefont {Zhang},\ and\ \citenamefont {Meng}}]{Lian_2018}%
  \BibitemOpen
  \bibfield  {author} {\bibinfo {author} {\bibfnamefont {C.}~\bibnamefont {Lian}}, \bibinfo {author} {\bibfnamefont {M.}~\bibnamefont {Guan}}, \bibinfo {author} {\bibfnamefont {S.}~\bibnamefont {Hu}}, \bibinfo {author} {\bibfnamefont {J.}~\bibnamefont {Zhang}},\ and\ \bibinfo {author} {\bibfnamefont {S.}~\bibnamefont {Meng}},\ }\bibfield  {title} {\bibinfo {title} {{Photoexcitation in Solids: First-Principles Quantum Simulations by Real-Time TDDFT}},\ }\href {https://doi.org/https://doi.org/10.1002/adts.201800055} {\bibfield  {journal} {\bibinfo  {journal} {Advanced Theory and Simulations}\ }\textbf {\bibinfo {volume} {1}},\ \bibinfo {pages} {1800055} (\bibinfo {year} {2018})}\BibitemShut {NoStop}%
\bibitem [{\citenamefont {Lipavsk\'y}\ \emph {et~al.}(1986)\citenamefont {Lipavsk\'y}, \citenamefont {\ifmmode \check{S}\else \v{S}\fi{}pi\ifmmode~\check{c}\else \v{c}\fi{}ka},\ and\ \citenamefont {Velick\'y}}]{Lipavsky_1986}%
  \BibitemOpen
  \bibfield  {author} {\bibinfo {author} {\bibfnamefont {P.}~\bibnamefont {Lipavsk\'y}}, \bibinfo {author} {\bibfnamefont {V.}~\bibnamefont {\ifmmode \check{S}\else \v{S}\fi{}pi\ifmmode~\check{c}\else \v{c}\fi{}ka}},\ and\ \bibinfo {author} {\bibfnamefont {B.}~\bibnamefont {Velick\'y}},\ }\bibfield  {title} {\bibinfo {title} {{{Generalized {K}adanoff-{B}aym} ansatz for deriving quantum transport equations}},\ }\href {https://doi.org/10.1103/PhysRevB.34.6933} {\bibfield  {journal} {\bibinfo  {journal} {Phys. Rev. B}\ }\textbf {\bibinfo {volume} {34}},\ \bibinfo {pages} {6933} (\bibinfo {year} {1986})}\BibitemShut {NoStop}%
\bibitem [{\citenamefont {Reeves}\ \emph {et~al.}(2023)\citenamefont {Reeves}, \citenamefont {Zhu}, \citenamefont {Yang},\ and\ \citenamefont {Vl\ifmmode~\check{c}\else \v{c}\fi{}ek}}]{Reeves_2023}%
  \BibitemOpen
  \bibfield  {author} {\bibinfo {author} {\bibfnamefont {C.~C.}\ \bibnamefont {Reeves}}, \bibinfo {author} {\bibfnamefont {Y.}~\bibnamefont {Zhu}}, \bibinfo {author} {\bibfnamefont {C.}~\bibnamefont {Yang}},\ and\ \bibinfo {author} {\bibfnamefont {V.~c.~v.}\ \bibnamefont {Vl\ifmmode~\check{c}\else \v{c}\fi{}ek}},\ }\bibfield  {title} {\bibinfo {title} {{Unimportance of memory for the time nonlocal components of the Kadanoff-Baym equations}},\ }\href {https://doi.org/10.1103/PhysRevB.108.115152} {\bibfield  {journal} {\bibinfo  {journal} {Phys. Rev. B}\ }\textbf {\bibinfo {volume} {108}},\ \bibinfo {pages} {115152} (\bibinfo {year} {2023})}\BibitemShut {NoStop}%
\bibitem [{\citenamefont {Schl\"unzen}\ \emph {et~al.}(2017)\citenamefont {Schl\"unzen}, \citenamefont {Joost}, \citenamefont {Heidrich-Meisner},\ and\ \citenamefont {Bonitz}}]{Schlunzen_2017}%
  \BibitemOpen
  \bibfield  {author} {\bibinfo {author} {\bibfnamefont {N.}~\bibnamefont {Schl\"unzen}}, \bibinfo {author} {\bibfnamefont {J.-P.}\ \bibnamefont {Joost}}, \bibinfo {author} {\bibfnamefont {F.}~\bibnamefont {Heidrich-Meisner}},\ and\ \bibinfo {author} {\bibfnamefont {M.}~\bibnamefont {Bonitz}},\ }\bibfield  {title} {\bibinfo {title} {{Nonequilibrium dynamics in the one-dimensional Fermi-Hubbard model: Comparison of the nonequilibrium Green-functions approach and the density matrix renormalization group method}},\ }\href {https://doi.org/10.1103/PhysRevB.95.165139} {\bibfield  {journal} {\bibinfo  {journal} {Phys. Rev. B}\ }\textbf {\bibinfo {volume} {95}},\ \bibinfo {pages} {165139} (\bibinfo {year} {2017})}\BibitemShut {NoStop}%
\bibitem [{\citenamefont {Bonitz}\ \emph {et~al.}(2013)\citenamefont {Bonitz}, \citenamefont {Hermanns},\ and\ \citenamefont {Balzer}}]{Bonitz_2013}%
  \BibitemOpen
  \bibfield  {author} {\bibinfo {author} {\bibfnamefont {M.}~\bibnamefont {Bonitz}}, \bibinfo {author} {\bibfnamefont {S.}~\bibnamefont {Hermanns}},\ and\ \bibinfo {author} {\bibfnamefont {K.}~\bibnamefont {Balzer}},\ }\bibfield  {title} {\bibinfo {title} {{Dynamics of {H}ubbard Nano-Clusters Following Strong Excitation}},\ }\href {https://doi.org/https://doi.org/10.1002/ctpp.201310053} {\bibfield  {journal} {\bibinfo  {journal} {Contributions to Plasma Physics}\ }\textbf {\bibinfo {volume} {53}},\ \bibinfo {pages} {778} (\bibinfo {year} {2013})}\BibitemShut {NoStop}%
\bibitem [{\citenamefont {Balzer}\ \emph {et~al.}(2013)\citenamefont {Balzer}, \citenamefont {Hermanns},\ and\ \citenamefont {Bonitz}}]{Balzer_2013}%
  \BibitemOpen
  \bibfield  {author} {\bibinfo {author} {\bibfnamefont {K.}~\bibnamefont {Balzer}}, \bibinfo {author} {\bibfnamefont {S.}~\bibnamefont {Hermanns}},\ and\ \bibinfo {author} {\bibfnamefont {M.}~\bibnamefont {Bonitz}},\ }\bibfield  {title} {\bibinfo {title} {{The generalized {K}adanoff-{B}aym ansatz. Computing nonlinear response properties of finite systems}},\ }\href {https://doi.org/10.1088/1742-6596/427/1/012006} {\bibfield  {journal} {\bibinfo  {journal} {Journal of Physics: Conference Series}\ }\textbf {\bibinfo {volume} {427}},\ \bibinfo {pages} {012006} (\bibinfo {year} {2013})}\BibitemShut {NoStop}%
\bibitem [{\citenamefont {Hermanns}\ \emph {et~al.}(2013)\citenamefont {Hermanns}, \citenamefont {Balzer},\ and\ \citenamefont {Bonitz}}]{Hermanns_2013}%
  \BibitemOpen
  \bibfield  {author} {\bibinfo {author} {\bibfnamefont {S.}~\bibnamefont {Hermanns}}, \bibinfo {author} {\bibfnamefont {K.}~\bibnamefont {Balzer}},\ and\ \bibinfo {author} {\bibfnamefont {M.}~\bibnamefont {Bonitz}},\ }\bibfield  {title} {\bibinfo {title} {{Few-particle quantum dynamics–comparing nonequilibrium {G}reen functions with the generalized {K}adanoff–{B}aym ansatz to density operator theory}},\ }\href {https://doi.org/10.1088/1742-6596/427/1/012008} {\bibfield  {journal} {\bibinfo  {journal} {Journal of Physics: Conference Series}\ }\textbf {\bibinfo {volume} {427}},\ \bibinfo {pages} {012008} (\bibinfo {year} {2013})}\BibitemShut {NoStop}%
\bibitem [{\citenamefont {Joost}\ \emph {et~al.}(2020)\citenamefont {Joost}, \citenamefont {Schl\"unzen},\ and\ \citenamefont {Bonitz}}]{Joost_2020}%
  \BibitemOpen
  \bibfield  {author} {\bibinfo {author} {\bibfnamefont {J.-P.}\ \bibnamefont {Joost}}, \bibinfo {author} {\bibfnamefont {N.}~\bibnamefont {Schl\"unzen}},\ and\ \bibinfo {author} {\bibfnamefont {M.}~\bibnamefont {Bonitz}},\ }\bibfield  {title} {\bibinfo {title} {{G1-G2 scheme: Dramatic acceleration of nonequilibrium Green functions simulations within the Hartree-Fock generalized Kadanoff-Baym ansatz}},\ }\href {https://doi.org/10.1103/PhysRevB.101.245101} {\bibfield  {journal} {\bibinfo  {journal} {Phys. Rev. B}\ }\textbf {\bibinfo {volume} {101}},\ \bibinfo {pages} {245101} (\bibinfo {year} {2020})}\BibitemShut {NoStop}%
\bibitem [{\citenamefont {von Friesen}\ \emph {et~al.}(2009)\citenamefont {von Friesen}, \citenamefont {Verdozzi},\ and\ \citenamefont {Almbladh}}]{von_Friesen_2009}%
  \BibitemOpen
  \bibfield  {author} {\bibinfo {author} {\bibfnamefont {M.~P.}\ \bibnamefont {von Friesen}}, \bibinfo {author} {\bibfnamefont {C.}~\bibnamefont {Verdozzi}},\ and\ \bibinfo {author} {\bibfnamefont {C.-O.}\ \bibnamefont {Almbladh}},\ }\bibfield  {title} {\bibinfo {title} {{Successes and Failures of {K}adanoff-{B}aym Dynamics in {H}ubbard Nanoclusters}},\ }\href {https://doi.org/10.1103/PhysRevLett.103.176404} {\bibfield  {journal} {\bibinfo  {journal} {Phys. Rev. Lett.}\ }\textbf {\bibinfo {volume} {103}},\ \bibinfo {pages} {176404} (\bibinfo {year} {2009})}\BibitemShut {NoStop}%
\bibitem [{\citenamefont {von Friesen}\ \emph {et~al.}(2010)\citenamefont {von Friesen}, \citenamefont {Verdozzi},\ and\ \citenamefont {Almbladh}}]{von_Friesen_2010}%
  \BibitemOpen
  \bibfield  {author} {\bibinfo {author} {\bibfnamefont {M.~P.}\ \bibnamefont {von Friesen}}, \bibinfo {author} {\bibfnamefont {C.}~\bibnamefont {Verdozzi}},\ and\ \bibinfo {author} {\bibfnamefont {C.-O.}\ \bibnamefont {Almbladh}},\ }\bibfield  {title} {\bibinfo {title} {{Artificial damping in the {K}adanoff-{B}aym dynamics of small {H}ubbard chains}},\ }\href {https://doi.org/10.1088/1742-6596/220/1/012016} {\bibfield  {journal} {\bibinfo  {journal} {Journal of Physics: Conference Series}\ }\textbf {\bibinfo {volume} {220}},\ \bibinfo {pages} {012016} (\bibinfo {year} {2010})}\BibitemShut {NoStop}%
\bibitem [{\citenamefont {Tuovinen}\ \emph {et~al.}(2020)\citenamefont {Tuovinen}, \citenamefont {Gole\ifmmode~\check{z}\else \v{z}\fi{}}, \citenamefont {Eckstein},\ and\ \citenamefont {Sentef}}]{Tuovinen_2020}%
  \BibitemOpen
  \bibfield  {author} {\bibinfo {author} {\bibfnamefont {R.}~\bibnamefont {Tuovinen}}, \bibinfo {author} {\bibfnamefont {D.}~\bibnamefont {Gole\ifmmode~\check{z}\else \v{z}\fi{}}}, \bibinfo {author} {\bibfnamefont {M.}~\bibnamefont {Eckstein}},\ and\ \bibinfo {author} {\bibfnamefont {M.~A.}\ \bibnamefont {Sentef}},\ }\bibfield  {title} {\bibinfo {title} {{Comparing the generalized Kadanoff-Baym ansatz with the full Kadanoff-Baym equations for an excitonic insulator out of equilibrium}},\ }\href {https://doi.org/10.1103/PhysRevB.102.115157} {\bibfield  {journal} {\bibinfo  {journal} {Phys. Rev. B}\ }\textbf {\bibinfo {volume} {102}},\ \bibinfo {pages} {115157} (\bibinfo {year} {2020})}\BibitemShut {NoStop}%
\bibitem [{sup()}]{supp}%
  \BibitemOpen
  \href@noop {} {}\bibinfo {note} {See Supplemental Material at \url{} for a selection of other results and discussions including: demonstrating the RT-DE scheme for ground state properties, long wavelength electric field excitation and system quench; derivations of the RT-DE equations of motion for $G^{\mathrm{R}}(t,t')$ and $G^{<}(t,t')$ for the second Born and $GW$ self-energies; a derivation showing the equivalence of the RT-DE to a one-shot correction on top of a mean-field reference; calculations showing TD-HF compared to exact results on the diagonal and implementation details}\BibitemShut {NoStop}%
\bibitem [{\citenamefont {Freericks}\ \emph {et~al.}(2009)\citenamefont {Freericks}, \citenamefont {Krishnamurthy},\ and\ \citenamefont {Pruschke}}]{Freericks_2009}%
  \BibitemOpen
  \bibfield  {author} {\bibinfo {author} {\bibfnamefont {J.~K.}\ \bibnamefont {Freericks}}, \bibinfo {author} {\bibfnamefont {H.~R.}\ \bibnamefont {Krishnamurthy}},\ and\ \bibinfo {author} {\bibfnamefont {T.}~\bibnamefont {Pruschke}},\ }\bibfield  {title} {\bibinfo {title} {{Theoretical Description of Time-Resolved Photoemission Spectroscopy: Application to Pump-Probe Experiments}},\ }\href {https://doi.org/10.1103/PhysRevLett.102.136401} {\bibfield  {journal} {\bibinfo  {journal} {Phys. Rev. Lett.}\ }\textbf {\bibinfo {volume} {102}},\ \bibinfo {pages} {136401} (\bibinfo {year} {2009})}\BibitemShut {NoStop}%
\bibitem [{\citenamefont {Boschini}\ \emph {et~al.}(2024)\citenamefont {Boschini}, \citenamefont {Zonno},\ and\ \citenamefont {Damascelli}}]{Boschini_2024}%
  \BibitemOpen
  \bibfield  {author} {\bibinfo {author} {\bibfnamefont {F.}~\bibnamefont {Boschini}}, \bibinfo {author} {\bibfnamefont {M.}~\bibnamefont {Zonno}},\ and\ \bibinfo {author} {\bibfnamefont {A.}~\bibnamefont {Damascelli}},\ }\bibfield  {title} {\bibinfo {title} {{Time-resolved ARPES studies of quantum materials}},\ }\href {https://doi.org/10.1103/RevModPhys.96.015003} {\bibfield  {journal} {\bibinfo  {journal} {Rev. Mod. Phys.}\ }\textbf {\bibinfo {volume} {96}},\ \bibinfo {pages} {015003} (\bibinfo {year} {2024})}\BibitemShut {NoStop}%
\bibitem [{\citenamefont {Choi}\ \emph {et~al.}(2023)\citenamefont {Choi}, \citenamefont {Yue}, \citenamefont {Azoury}, \citenamefont {Porter}, \citenamefont {Chen}, \citenamefont {Petocchi}, \citenamefont {Baldini}, \citenamefont {Lv}, \citenamefont {Mogi}, \citenamefont {Su}, \citenamefont {Wilson}, \citenamefont {Eckstein}, \citenamefont {Werner},\ and\ \citenamefont {Gedik}}]{choi_2023}%
  \BibitemOpen
  \bibfield  {author} {\bibinfo {author} {\bibfnamefont {D.}~\bibnamefont {Choi}}, \bibinfo {author} {\bibfnamefont {C.}~\bibnamefont {Yue}}, \bibinfo {author} {\bibfnamefont {D.}~\bibnamefont {Azoury}}, \bibinfo {author} {\bibfnamefont {Z.}~\bibnamefont {Porter}}, \bibinfo {author} {\bibfnamefont {J.}~\bibnamefont {Chen}}, \bibinfo {author} {\bibfnamefont {F.}~\bibnamefont {Petocchi}}, \bibinfo {author} {\bibfnamefont {E.}~\bibnamefont {Baldini}}, \bibinfo {author} {\bibfnamefont {B.}~\bibnamefont {Lv}}, \bibinfo {author} {\bibfnamefont {M.}~\bibnamefont {Mogi}}, \bibinfo {author} {\bibfnamefont {Y.}~\bibnamefont {Su}}, \bibinfo {author} {\bibfnamefont {S.~D.}\ \bibnamefont {Wilson}}, \bibinfo {author} {\bibfnamefont {M.}~\bibnamefont {Eckstein}}, \bibinfo {author} {\bibfnamefont {P.}~\bibnamefont {Werner}},\ and\ \bibinfo {author} {\bibfnamefont {N.}~\bibnamefont {Gedik}},\ }\href@noop {} {\bibinfo {title} {{Light-induced insulator-metal transition in Sr$_2$IrO$_4$ reveals the nature of the insulating
  ground state}}} (\bibinfo {year} {2023}),\ \Eprint {https://arxiv.org/abs/2305.07619} {arXiv:2305.07619 [cond-mat.str-el]} \BibitemShut {NoStop}%
\bibitem [{\citenamefont {Latini}\ \emph {et~al.}(2014)\citenamefont {Latini}, \citenamefont {Perfetto}, \citenamefont {Uimonen}, \citenamefont {van Leeuwen},\ and\ \citenamefont {Stefanucci}}]{Latini_2014}%
  \BibitemOpen
  \bibfield  {author} {\bibinfo {author} {\bibfnamefont {S.}~\bibnamefont {Latini}}, \bibinfo {author} {\bibfnamefont {E.}~\bibnamefont {Perfetto}}, \bibinfo {author} {\bibfnamefont {A.-M.}\ \bibnamefont {Uimonen}}, \bibinfo {author} {\bibfnamefont {R.}~\bibnamefont {van Leeuwen}},\ and\ \bibinfo {author} {\bibfnamefont {G.}~\bibnamefont {Stefanucci}},\ }\bibfield  {title} {\bibinfo {title} {{Charge dynamics in molecular junctions: Nonequilibrium Green's function approach made fast}},\ }\href {https://doi.org/10.1103/PhysRevB.89.075306} {\bibfield  {journal} {\bibinfo  {journal} {Phys. Rev. B}\ }\textbf {\bibinfo {volume} {89}},\ \bibinfo {pages} {075306} (\bibinfo {year} {2014})}\BibitemShut {NoStop}%
\bibitem [{\citenamefont {Hopjan}\ and\ \citenamefont {Verdozzi}(2019)}]{Hopjan_2019}%
  \BibitemOpen
  \bibfield  {author} {\bibinfo {author} {\bibfnamefont {M.}~\bibnamefont {Hopjan}}\ and\ \bibinfo {author} {\bibfnamefont {C.}~\bibnamefont {Verdozzi}},\ }\bibfield  {title} {\bibinfo {title} {{Initial correlated states for the generalized Kadanoff–Baym Ansatz without adiabatic switching-on of interactions in closed systems}},\ }\href {https://doi.org/10.1140/epjst/e2018-800054-3} {\bibfield  {journal} {\bibinfo  {journal} {The European Physical Journal Special Topics}\ }\textbf {\bibinfo {volume} {227}},\ \bibinfo {pages} {1939–1948} (\bibinfo {year} {2019})}\BibitemShut {NoStop}%
\bibitem [{\citenamefont {Sch\"uler}\ \emph {et~al.}(2020)\citenamefont {Sch\"uler}, \citenamefont {De~Giovannini}, \citenamefont {H\"ubener}, \citenamefont {Rubio}, \citenamefont {Sentef}, \citenamefont {Devereaux},\ and\ \citenamefont {Werner}}]{Schuler_2020}%
  \BibitemOpen
  \bibfield  {author} {\bibinfo {author} {\bibfnamefont {M.}~\bibnamefont {Sch\"uler}}, \bibinfo {author} {\bibfnamefont {U.}~\bibnamefont {De~Giovannini}}, \bibinfo {author} {\bibfnamefont {H.}~\bibnamefont {H\"ubener}}, \bibinfo {author} {\bibfnamefont {A.}~\bibnamefont {Rubio}}, \bibinfo {author} {\bibfnamefont {M.~A.}\ \bibnamefont {Sentef}}, \bibinfo {author} {\bibfnamefont {T.~P.}\ \bibnamefont {Devereaux}},\ and\ \bibinfo {author} {\bibfnamefont {P.}~\bibnamefont {Werner}},\ }\bibfield  {title} {\bibinfo {title} {{How Circular Dichroism in Time- and Angle-Resolved Photoemission Can Be Used to Spectroscopically Detect Transient Topological States in Graphene}},\ }\href {https://doi.org/10.1103/PhysRevX.10.041013} {\bibfield  {journal} {\bibinfo  {journal} {Phys. Rev. X}\ }\textbf {\bibinfo {volume} {10}},\ \bibinfo {pages} {041013} (\bibinfo {year} {2020})}\BibitemShut {NoStop}%
\bibitem [{\citenamefont {Stan}\ \emph {et~al.}(2009)\citenamefont {Stan}, \citenamefont {Dahlen},\ and\ \citenamefont {van Leeuwen}}]{Stan_2009}%
  \BibitemOpen
  \bibfield  {author} {\bibinfo {author} {\bibfnamefont {A.}~\bibnamefont {Stan}}, \bibinfo {author} {\bibfnamefont {N.}~\bibnamefont {Dahlen}},\ and\ \bibinfo {author} {\bibfnamefont {R.}~\bibnamefont {van Leeuwen}},\ }\bibfield  {title} {\bibinfo {title} {{Time propagation of the {K}adanoff{\textendash}{B}aym equations for inhomogeneous systems}},\ }\href {https://doi.org/10.1063/1.3127247} {\bibfield  {journal} {\bibinfo  {journal} {J. Chem. Phys.}\ }\textbf {\bibinfo {volume} {130}},\ \bibinfo {pages} {224101} (\bibinfo {year} {2009})}\BibitemShut {NoStop}%
\bibitem [{\citenamefont {Stefanucci}\ and\ \citenamefont {van Leeuwen}(2013)}]{stefanucci2013nonequilibrium}%
  \BibitemOpen
  \bibfield  {author} {\bibinfo {author} {\bibfnamefont {G.}~\bibnamefont {Stefanucci}}\ and\ \bibinfo {author} {\bibfnamefont {R.}~\bibnamefont {van Leeuwen}},\ }\href@noop {} {\emph {\bibinfo {title} {{Nonequilibrium Many-Body Theory of Quantum Systems: A Modern Introduction}}}}\ (\bibinfo  {publisher} {Cambridge University Press},\ \bibinfo {year} {2013})\BibitemShut {NoStop}%
\bibitem [{\citenamefont {Yin}\ \emph {et~al.}(2022)\citenamefont {Yin}, \citenamefont {hao Chan}, \citenamefont {da~Jornada}, \citenamefont {Qiu}, \citenamefont {Louie},\ and\ \citenamefont {Yang}}]{Yin_2022}%
  \BibitemOpen
  \bibfield  {author} {\bibinfo {author} {\bibfnamefont {J.}~\bibnamefont {Yin}}, \bibinfo {author} {\bibfnamefont {Y.}~\bibnamefont {hao Chan}}, \bibinfo {author} {\bibfnamefont {F.~H.}\ \bibnamefont {da~Jornada}}, \bibinfo {author} {\bibfnamefont {D.~Y.}\ \bibnamefont {Qiu}}, \bibinfo {author} {\bibfnamefont {S.~G.}\ \bibnamefont {Louie}},\ and\ \bibinfo {author} {\bibfnamefont {C.}~\bibnamefont {Yang}},\ }\bibfield  {title} {\bibinfo {title} {{Using dynamic mode decomposition to predict the dynamics of a two-time non-equilibrium Green’s function}},\ }\href {https://doi.org/https://doi.org/10.1016/j.jocs.2022.101843} {\bibfield  {journal} {\bibinfo  {journal} {Journal of Computational Science}\ }\textbf {\bibinfo {volume} {64}},\ \bibinfo {pages} {101843} (\bibinfo {year} {2022})}\BibitemShut {NoStop}%
\bibitem [{\citenamefont {Kaye}\ and\ \citenamefont {U.~R.~Strand}(2023)}]{Kaye_2023}%
  \BibitemOpen
  \bibfield  {author} {\bibinfo {author} {\bibfnamefont {J.}~\bibnamefont {Kaye}}\ and\ \bibinfo {author} {\bibfnamefont {H.}~\bibnamefont {U.~R.~Strand}},\ }\bibfield  {title} {\bibinfo {title} {{A fast time domain solver for the equilibrium Dyson equation}},\ }\bibfield  {journal} {\bibinfo  {journal} {Advances in Computational Mathematics}\ }\textbf {\bibinfo {volume} {49}},\ \href {https://doi.org/10.1007/s10444-023-10067-7} {10.1007/s10444-023-10067-7} (\bibinfo {year} {2023})\BibitemShut {NoStop}%
\bibitem [{\citenamefont {Stahl}\ \emph {et~al.}(2022)\citenamefont {Stahl}, \citenamefont {Dasari}, \citenamefont {Li}, \citenamefont {Picano}, \citenamefont {Werner},\ and\ \citenamefont {Eckstein}}]{Stahl_2022}%
  \BibitemOpen
  \bibfield  {author} {\bibinfo {author} {\bibfnamefont {C.}~\bibnamefont {Stahl}}, \bibinfo {author} {\bibfnamefont {N.}~\bibnamefont {Dasari}}, \bibinfo {author} {\bibfnamefont {J.}~\bibnamefont {Li}}, \bibinfo {author} {\bibfnamefont {A.}~\bibnamefont {Picano}}, \bibinfo {author} {\bibfnamefont {P.}~\bibnamefont {Werner}},\ and\ \bibinfo {author} {\bibfnamefont {M.}~\bibnamefont {Eckstein}},\ }\bibfield  {title} {\bibinfo {title} {{Memory truncated Kadanoff-Baym equations}},\ }\href {https://doi.org/10.1103/PhysRevB.105.115146} {\bibfield  {journal} {\bibinfo  {journal} {Phys. Rev. B}\ }\textbf {\bibinfo {volume} {105}},\ \bibinfo {pages} {115146} (\bibinfo {year} {2022})}\BibitemShut {NoStop}%
\bibitem [{\citenamefont {Blommel}\ \emph {et~al.}(2024)\citenamefont {Blommel}, \citenamefont {Kaye}, \citenamefont {Murakami}, \citenamefont {Gull},\ and\ \citenamefont {Golež}}]{Blommel_2024}%
  \BibitemOpen
  \bibfield  {author} {\bibinfo {author} {\bibfnamefont {T.}~\bibnamefont {Blommel}}, \bibinfo {author} {\bibfnamefont {J.}~\bibnamefont {Kaye}}, \bibinfo {author} {\bibfnamefont {Y.}~\bibnamefont {Murakami}}, \bibinfo {author} {\bibfnamefont {E.}~\bibnamefont {Gull}},\ and\ \bibinfo {author} {\bibfnamefont {D.}~\bibnamefont {Golež}},\ }\href@noop {} {\bibinfo {title} {{Chirped amplitude mode in photo-excited superconductors}}} (\bibinfo {year} {2024}),\ \Eprint {https://arxiv.org/abs/2403.01589} {arXiv:2403.01589 [cond-mat.supr-con]} \BibitemShut {NoStop}%
\bibitem [{\citenamefont {Randi}\ \emph {et~al.}(2017)\citenamefont {Randi}, \citenamefont {Fausti},\ and\ \citenamefont {Eckstein}}]{Randi_2017}%
  \BibitemOpen
  \bibfield  {author} {\bibinfo {author} {\bibfnamefont {F.}~\bibnamefont {Randi}}, \bibinfo {author} {\bibfnamefont {D.}~\bibnamefont {Fausti}},\ and\ \bibinfo {author} {\bibfnamefont {M.}~\bibnamefont {Eckstein}},\ }\bibfield  {title} {\bibinfo {title} {{Bypassing the energy-time uncertainty in time-resolved photoemission}},\ }\href {https://doi.org/10.1103/PhysRevB.95.115132} {\bibfield  {journal} {\bibinfo  {journal} {Phys. Rev. B}\ }\textbf {\bibinfo {volume} {95}},\ \bibinfo {pages} {115132} (\bibinfo {year} {2017})}\BibitemShut {NoStop}%
\bibitem [{\citenamefont {Vacondio}\ \emph {et~al.}(2022)\citenamefont {Vacondio}, \citenamefont {Varsano}, \citenamefont {Ruini},\ and\ \citenamefont {Ferretti}}]{Vacondio_2022}%
  \BibitemOpen
  \bibfield  {author} {\bibinfo {author} {\bibfnamefont {S.}~\bibnamefont {Vacondio}}, \bibinfo {author} {\bibfnamefont {D.}~\bibnamefont {Varsano}}, \bibinfo {author} {\bibfnamefont {A.}~\bibnamefont {Ruini}},\ and\ \bibinfo {author} {\bibfnamefont {A.}~\bibnamefont {Ferretti}},\ }\bibfield  {title} {\bibinfo {title} {{Numerically Precise Benchmark of Many-Body Self-Energies on Spherical Atoms}},\ }\href {https://doi.org/10.1021/acs.jctc.2c00048} {\bibfield  {journal} {\bibinfo  {journal} {Journal of Chemical Theory and Computation}\ }\textbf {\bibinfo {volume} {18}},\ \bibinfo {pages} {3703} (\bibinfo {year} {2022})}\BibitemShut {NoStop}%
\bibitem [{\citenamefont {Mejuto-Zaera}\ and\ \citenamefont {Vlček}(2022)}]{Mejuto_2022}%
  \BibitemOpen
  \bibfield  {author} {\bibinfo {author} {\bibfnamefont {C.}~\bibnamefont {Mejuto-Zaera}}\ and\ \bibinfo {author} {\bibfnamefont {V.}~\bibnamefont {Vlček}},\ }\bibfield  {title} {\bibinfo {title} {{Self-consistency in $GW\mathrm{\ensuremath{\Gamma}}$ formalism leading to quasiparticle-quasiparticle couplings}},\ }\href {https://doi.org/10.1103/PhysRevB.106.165129} {\bibfield  {journal} {\bibinfo  {journal} {Phys. Rev. B}\ }\textbf {\bibinfo {volume} {106}},\ \bibinfo {pages} {165129} (\bibinfo {year} {2022})}\BibitemShut {NoStop}%
\bibitem [{\citenamefont {Mejuto-Zaera}\ \emph {et~al.}(2021)\citenamefont {Mejuto-Zaera}, \citenamefont {Weng}, \citenamefont {Romanova}, \citenamefont {Cotton}, \citenamefont {Whaley}, \citenamefont {Tubman},\ and\ \citenamefont {Vlček}}]{Mejuto_2021}%
  \BibitemOpen
  \bibfield  {author} {\bibinfo {author} {\bibfnamefont {C.}~\bibnamefont {Mejuto-Zaera}}, \bibinfo {author} {\bibfnamefont {G.}~\bibnamefont {Weng}}, \bibinfo {author} {\bibfnamefont {M.}~\bibnamefont {Romanova}}, \bibinfo {author} {\bibfnamefont {S.~J.}\ \bibnamefont {Cotton}}, \bibinfo {author} {\bibfnamefont {K.~B.}\ \bibnamefont {Whaley}}, \bibinfo {author} {\bibfnamefont {N.~M.}\ \bibnamefont {Tubman}},\ and\ \bibinfo {author} {\bibfnamefont {V.}~\bibnamefont {Vlček}},\ }\bibfield  {title} {\bibinfo {title} {{{Are multi-quasiparticle interactions important in molecular ionization?}}},\ }\href {https://doi.org/10.1063/5.0044060} {\bibfield  {journal} {\bibinfo  {journal} {The Journal of Chemical Physics}\ }\textbf {\bibinfo {volume} {154}},\ \bibinfo {pages} {121101} (\bibinfo {year} {2021})}\BibitemShut {NoStop}%
\bibitem [{\citenamefont {Vlček}(2019)}]{Vlcek_2019}%
  \BibitemOpen
  \bibfield  {author} {\bibinfo {author} {\bibfnamefont {V.}~\bibnamefont {Vlček}},\ }\bibfield  {title} {\bibinfo {title} {{Stochastic Vertex Corrections: Linear Scaling Methods for Accurate Quasiparticle Energies}},\ }\href {https://doi.org/10.1021/acs.jctc.9b00317} {\bibfield  {journal} {\bibinfo  {journal} {Journal of Chemical Theory and Computation}\ }\textbf {\bibinfo {volume} {15}},\ \bibinfo {pages} {6254} (\bibinfo {year} {2019})},\ \bibinfo {note} {pMID: 31557012}\BibitemShut {NoStop}%
\bibitem [{\citenamefont {Ullrich}(2012)}]{ullrich_2012}%
  \BibitemOpen
  \bibfield  {author} {\bibinfo {author} {\bibfnamefont {C.}~\bibnamefont {Ullrich}},\ }\href {https://books.google.com/books?id=hCNNsC4sEtkC} {\emph {\bibinfo {title} {Time-Dependent Density-Functional Theory: Concepts and Applications}}},\ Oxford Graduate Texts\ (\bibinfo  {publisher} {OUP Oxford,Great Clarendon St, Oxford},\ \bibinfo {year} {2012})\BibitemShut {NoStop}%
\end{thebibliography}%


\begin{thebibliography}{2}%
\makeatletter
\providecommand \@ifxundefined [1]{%
 \@ifx{#1\undefined}
}%
\providecommand \@ifnum [1]{%
 \ifnum #1\expandafter \@firstoftwo
 \else \expandafter \@secondoftwo
 \fi
}%
\providecommand \@ifx [1]{%
 \ifx #1\expandafter \@firstoftwo
 \else \expandafter \@secondoftwo
 \fi
}%
\providecommand \natexlab [1]{#1}%
\providecommand \enquote  [1]{``#1''}%
\providecommand \bibnamefont  [1]{#1}%
\providecommand \bibfnamefont [1]{#1}%
\providecommand \citenamefont [1]{#1}%
\providecommand \href@noop [0]{\@secondoftwo}%
\providecommand \href [0]{\begingroup \@sanitize@url \@href}%
\providecommand \@href[1]{\@@startlink{#1}\@@href}%
\providecommand \@@href[1]{\endgroup#1\@@endlink}%
\providecommand \@sanitize@url [0]{\catcode `\\12\catcode `\$12\catcode `\&12\catcode `\#12\catcode `\^12\catcode `\_12\catcode `\%12\relax}%
\providecommand \@@startlink[1]{}%
\providecommand \@@endlink[0]{}%
\providecommand \url  [0]{\begingroup\@sanitize@url \@url }%
\providecommand \@url [1]{\endgroup\@href {#1}{\urlprefix }}%
\providecommand \urlprefix  [0]{URL }%
\providecommand \Eprint [0]{\href }%
\providecommand \doibase [0]{https://doi.org/}%
\providecommand \selectlanguage [0]{\@gobble}%
\providecommand \bibinfo  [0]{\@secondoftwo}%
\providecommand \bibfield  [0]{\@secondoftwo}%
\providecommand \translation [1]{[#1]}%
\providecommand \BibitemOpen [0]{}%
\providecommand \bibitemStop [0]{}%
\providecommand \bibitemNoStop [0]{.\EOS\space}%
\providecommand \EOS [0]{\spacefactor3000\relax}%
\providecommand \BibitemShut  [1]{\csname bibitem#1\endcsname}%
\let\auto@bib@innerbib\@empty
\bibitem [{\citenamefont {Ullrich}(2012)}]{ullrich_2012}%
  \BibitemOpen
  \bibfield  {author} {\bibinfo {author} {\bibfnamefont {C.}~\bibnamefont {Ullrich}},\ }\href {https://books.google.com/books?id=hCNNsC4sEtkC} {\emph {\bibinfo {title} {Time-Dependent Density-Functional Theory: Concepts and Applications}}},\ Oxford Graduate Texts\ (\bibinfo  {publisher} {OUP Oxford,Great Clarendon St, Oxford},\ \bibinfo {year} {2012})\BibitemShut {NoStop}%
\bibitem [{\citenamefont {Joost}\ \emph {et~al.}(2020)\citenamefont {Joost}, \citenamefont {Schl\"unzen},\ and\ \citenamefont {Bonitz}}]{Joost_2020}%
  \BibitemOpen
  \bibfield  {author} {\bibinfo {author} {\bibfnamefont {J.-P.}\ \bibnamefont {Joost}}, \bibinfo {author} {\bibfnamefont {N.}~\bibnamefont {Schl\"unzen}},\ and\ \bibinfo {author} {\bibfnamefont {M.}~\bibnamefont {Bonitz}},\ }\bibfield  {title} {\bibinfo {title} {{G1-G2 scheme: Dramatic acceleration of nonequilibrium Green functions simulations within the Hartree-Fock generalized Kadanoff-Baym ansatz}},\ }\href {https://doi.org/10.1103/PhysRevB.101.245101} {\bibfield  {journal} {\bibinfo  {journal} {Phys. Rev. B}\ }\textbf {\bibinfo {volume} {101}},\ \bibinfo {pages} {245101} (\bibinfo {year} {2020})}\BibitemShut {NoStop}%
\end{thebibliography}%
\end{document}


\preprint{APS/123-QED}

\title{Supplementary information for: ``A Real-time Dyson Expansion Scheme: Efficient Inclusion of Dynamical Correlations in Non-equilibrium Spectral Properties"}
\author{Cian C. Reeves}
\affiliation{%
Department of Physics, University of California, Santa Barbara, Santa Barbara, CA 93117
}%
\author{Vojt\ifmmode \check{e}\else \v{e}ch Vl\v{c}ek}
\affiliation{%
Department of Chemistry and Biochemistry, University of California, Santa Barbara, Santa Barbara, CA 93117
}%
\affiliation{%
Department of Materials, University of California, Santa Barbara, Santa Barbara, CA 93117
}

\date{\today}

\maketitle

\section{Derivation of the RT-DE Equations of Motion with the second Born self-energy}\label{sec:SB_derivation}
We will now derive the equations of motion for the RT-DE ODE scheme using the second Born self-energy.  We start by deriving the equations for the retarded component of the Green's function.  The equation of motion for $G^{\mathrm{R}}(t,t')$ is given by
\begin{equation*}
\begin{split}
        \left[i\partial_t - h^{\mathrm{MF}}(t)\right]G^{\mathrm{R}}(t,t') &= \delta(t,t') + \int_{t'}^t d\Bar{t}\medspace \Sigma^{\mathrm{R}}(t,\Bar{t}) G^{\mathrm{R}}(\Bar{t},t')\\
        &= \delta(t,t') + I^{R}(t,t')\\
\end{split}
\end{equation*}
For the second Born self-energy $\Sigma^{\mathrm{R}}$ is given by
    
\begin{equation*}
    \begin{split}
        \Sigma^{\mathrm{R}}_{ij}(t,t') &= -\sum_{klpqrs} w_{iklp}(t)w_{qrsj}(t')\bigg{[}G^>_{lq}(t,t')G^>_{pr}(t,t')G^<_{sk}(t',t) 
        - G^<_{lq}(t,t')G^<_{pr}(t,t')G^>_{sk}(t',t)\bigg{]},
    \end{split}
\end{equation*}

we can write,
\begin{equation*}
\begin{split}
        I^{\mathrm{R}}_{im}(t,t') &= -\sum_{klp} w_{iklp}(t)\bigg{[}\sum_{qrsj}\int_{t'}^t d\Bar{t}\medspace w_{qrsj}(\Bar{t})\Big{[}G^>_{lq}(t,\Bar{t})G^>_{pr}(t,\Bar{t})G^<_{sk}(\Bar{t},t) - G^<_{lq}(t,\Bar{t})G^<_{pr}(t,\Bar{t})G^>_{sk}(\Bar{t},t)\Big{]}G^{\mathrm{R}}_{jm}(\Bar{t},t')\bigg{]} \\
        &= -\sum_{klp}w_{iklp}(t) \mathcal{F}_{lpmk}(t,t').
\end{split}
\end{equation*}

At this point we make the approximation that $\Sigma^{\mathrm{R}}(t,t')$ constructed from mean-field propagators.  This amounts to making the following replacement for $G^\gtrless(t,t')$,

\begin{equation}\label{eq:G_MF}
    \begin{split}
        G^{\gtrless,\mathrm{MF}}_{ij}(t' \leq t) &= i \sum_{k}G^{\gtrless,\mathrm{MF}}_{ik}(t',t')(t)\mathcal{U}_{kj}(t',t),\\
        G^{\gtrless,\mathrm{MF}}_{ij}(t\geq t') &= i \sum_{k}\mathcal{U}_{ik}(t,t')G^{\gtrless,\mathrm{MF}}_{kj}(t',t').\\
    \end{split}
\end{equation}
Here $\mathcal{U}(t,t')$ satisfies,
\begin{equation}\label{eq:U_eom}
   \begin{split}
        \frac{\mathrm{d}\mathcal{U}_{ij}(t,t')}{\mathrm{d}t} &= -i\sum_k h^{\mathrm{MF}}_{ik}(t)\mathcal{U}_{kj}(t,t'),\\
         \frac{\mathrm{d}\mathcal{U}_{ij}(t',t)}{\mathrm{d}t} &= i\sum_k\mathcal{U}_{ik}(t',t)h^{\mathrm{MF}}_{kj}(t),\\
         \mathcal{U}_{ij}(t,t) &= \delta_{ij}.
   \end{split}
\end{equation}

Taking this approximation into account we can rewrite $\mathcal{F}(t,t')$ as
\begin{equation}
    \begin{split}
        \mathcal{F}_{lpmk}(t,t') &= \sum_{qrsjzno}\int_{t'}^t d\Bar{t}\medspace w_{qrsj}(\Bar{t})\mathcal{U}_{lz}(t,\Bar{t}) \mathcal{U}_{pn}(t,\Bar{t})\bigg{[}G^{>,\mathrm{MF}}_{zq}(\Bar{t})G^{>,\mathrm{MF}}_{nr}(\Bar{t})G^{<,\mathrm{MF}}_{so}(\Bar{t}) \\&\hspace{60mm}- G^{<,\mathrm{MF}}_{zq}(\Bar{t})G^{<,\mathrm{MF}}_{nr}(\Bar{t})G^{>,\mathrm{MF}}_{so}(\Bar{t})\bigg{]}\mathcal{U}_{ok}(\Bar{t},t)G^{\mathrm{R}}_{jm}(\Bar{t},t').
    \end{split}
\end{equation}
With the equation of motion for $\mathcal{U}(t,t')$ we can derive an ODE equation for $\mathcal{F}_{lkpm}(t,t')$.  First, taking the time derivative of the integral we have,
\begin{equation*}
 \begin{split}
    \left[\frac{\mathrm{d}\mathcal{F}_{lpmk}(t,t')}{\mathrm{d}t}\right]_{\int} &= \sum_{qrsj} w_{qrsj}(t)\Big[ G^{>,\mathrm{MF}}_{lq}(t)G^{>,\mathrm{MF}}_{pr}(t)G^{<,\mathrm{MF}}_{sk}(t) - G^{<,\mathrm{MF}}_{lq}(t)G^{<,\mathrm{MF}}_{pr}(t)G^{>,\mathrm{MF}}_{sk}(t)\Big{]} G^{\mathrm{R}}_{jm}(t,t').
 \end{split}
\end{equation*}
Next we take the derivative involving the $\mathcal{U}(t,t')$ operators.  Using the equations of motion for $\mathcal{U}(t,t')$ in equation \eqref{eq:U_eom} we have
\begin{equation*}
    \begin{split}
&\left[\frac{\mathrm{d}\mathcal{F}_{lpmk}(t,t')}{\mathrm{d}t}\right]_{\mathcal{U}} 
\\&=-i\sum_{qrsjznox}\int_{t'}^t d\Bar{t}\medspace w_{qrsj}(\Bar{t})\bigg{\{}\\&h^{\mathrm{MF}}_{lx}(t)\mathcal{U}_{xz}(t,\Bar{t}) \mathcal{U}_{pn}(t,\Bar{t})\Big{[}G^{>,\mathrm{MF}}_{zq}(\Bar{t})G^{>,\mathrm{MF}}_{nr}(\Bar{t})G^{<,\mathrm{MF}}_{so}(\Bar{t}) - G^{<,\mathrm{MF}}_{zq}(\Bar{t})G^{<,\mathrm{MF}}_{nr}(\Bar{t})G^{>,\mathrm{MF}}_{so}(\Bar{t})\Big{]}\mathcal{U}_{ok}(\Bar{t},t)G^{\mathrm{R}}_{jm}(\Bar{t},t')
        \\&+h^{\mathrm{MF}}_{px}(t)\mathcal{U}_{lz}(t,\Bar{t}) \mathcal{U}_{xn}(t,\Bar{t})\Big{[}G^{>,\mathrm{MF}}_{zq}(\Bar{t})G^{>,\mathrm{MF}}_{nr}(\Bar{t})G^{<,\mathrm{MF}}_{so}(\Bar{t}) - G^{<,\mathrm{MF}}_{zq}(\Bar{t})G^{<,\mathrm{MF}}_{nr}(\Bar{t})G^{>,\mathrm{MF}}_{so}(\Bar{t})\Big{]}\mathcal{U}_{ok}(\Bar{t},t)G^{\mathrm{R}}_{jm}(\Bar{t},t')\\&-\mathcal{U}_{lz}(t,\Bar{t}) \mathcal{U}_{pn}(t,\Bar{t})\Big{[}G^{>,\mathrm{MF}}_{zq}(\Bar{t})G^{>,\mathrm{MF}}_{nr}(\Bar{t})G^{<,\mathrm{MF}}_{so}(\Bar{t}) - G^{<,\mathrm{MF}}_{zq}(\Bar{t})G^{<,\mathrm{MF}}_{nr}(\Bar{t})G^{>,\mathrm{MF}}_{so}(\Bar{t})\Big{]}\mathcal{U}_{ox(\Bar{t},t)}h^{\mathrm{MF}}_{xk}(t)G^{\mathrm{R}}_{jm}(\Bar{t},t')\bigg{\}},\\
        &= -i\sum_{x}\Big{[} h^\mathrm{MF}_{lx}(t) \mathcal{F}_{xpmk}(t,t') + h^\mathrm{MF}_{px}(t) \mathcal{F}_{lxmk} (t,t')-   \mathcal{F}_{lpmx}(t,t') h^\mathrm{MF}_{xk}(t)\Big{]}.
    \end{split}
\end{equation*}
We are now left with the following set of equations,
\begin{equation}\label{RT-DE}
       \begin{split}
       \frac{\mathrm{d}G^{\mathrm{R}}(t,t')}{\mathrm{d}t} &= -i\left[h^{\mathrm{MF}}(t) G^{\mathrm{R}}(t,t') + I^{\mathrm{R}}(t,t')\right],\\
       I^{\mathrm{R}}_{im}(t,t') &= -\sum_{klp}w_{iklp}(t) \mathcal{F}_{lpmk}(t,t'),\\
        \frac{\mathrm{d}\mathcal{F}_{lpmk}(t,t')}{\mathrm{d}t} &=  \sum_{qrsj} w_{qrsj}(t)\Big{[} G^{>,\mathrm{MF}}_{lq}(t)G^{>,\mathrm{MF}}_{pr}(t)G^{<,\mathrm{MF}}_{sk}(t) - G^{<,\mathrm{MF}}_{lq}(t)G^{<,\mathrm{MF}}_{pr}(t)G^{>,\mathrm{MF}}_{sk}(t)\Big{]} G^{\mathrm{R}}_{jm}(t,t') \\&\hspace{15mm}-i\sum_{x}\Big{[} h^\mathrm{MF}_{lx}(t) \mathcal{F}_{xpmk}(t,t') + h^\mathrm{MF}_{px}(t) \mathcal{F}_{lxmk} (t,t') - \mathcal{F}_{lpmx}(t,t') h^\mathrm{MF}_{xk}(t)\Big{]}.\\
   \end{split}
\end{equation}

The RT-DE scheme works by first performing a time evolution of $G(t,t)$.  Using the time diagonal components in equation \eqref{RT-DE} for each value of $t'$ we can time-step in the $t$ variable in the range $t'<t<T_{\mathrm{max}}$. We have the following initial conditions for each $t'$,
\begin{equation*}
\begin{split}
        G^{\mathrm{R}}(t',t') &= -i,\\
        \mathcal{F}(t',t') &= 0.
\end{split}
\end{equation*}
For the lesser component we have the following equation of motion,
\begin{equation}
    \begin{split}
        \left[i\partial_t - h^{\mathrm{MF}}(t)\right]G^{<}(t,t') &= \int_{0}^t d\Bar{t}\medspace\Sigma^{\mathrm{R}}(t,\Bar{t})G^{<}(\Bar{t},t') + \int_0^{t'}d\Bar{t}\medspace\Sigma^{>}(t,\Bar{t})G^{\mathrm{A}}(\Bar{t},t'),\\
        &= I^{<}(t,t')\\
        I_{im}^<(t,t') &= -\sum_{klp}w_{iklp}(t) \Tilde{\mathcal{F}}_{lpmk}(t,t')
    \end{split}
\end{equation}
Following the same steps as for the retarded component we have,
\begin{equation*}
\left[\frac{\mathrm{d}\Tilde{\mathcal{F}}_{lpmk}(t,t')}{\mathrm{d}t}\right]_{\int} = \sum_{qrsj} w_{qrsj}(t)\Big[ G^{>,\mathrm{MF}}_{lq}(t)G^{>,\mathrm{MF}}_{pr}(t)G^{<,\mathrm{MF}}_{sk}(t) - G^{<,\mathrm{MF}}_{lq}(t)G^{<,\mathrm{MF}}_{pr}(t)G^{>,\mathrm{MF}}_{sk}(t)\Big{]} G^{<}_{jm}(t,t').
\end{equation*}
Similarly for the derivative with respect to $\mathcal{U}(t,t')$ we have,
\begin{equation*}
    \begin{split}
\left[\frac{\mathrm{d}\Tilde{\mathcal{F}}_{lpmk}(t,t')}{\mathrm{d}t}\right]_{\mathcal{U}} 
        &= -i\sum_{x}\Big{[} h^\mathrm{MF}_{lx}(t) \Tilde{\mathcal{F}}_{xpmk}(t,t') + h^\mathrm{MF}_{px}(t) \Tilde{\mathcal{F}}_{lxmk} (t,t')-  \Tilde{\mathcal{F}}_{lpmx}(t,t') h^\mathrm{MF}_{xk}(t)\Big{]}.
    \end{split}
\end{equation*}

Combining these we get a similar set of equations to equation \eqref{RT-DE} for the lesser component of the Green's function.
\begin{equation}\label{RT-DE_lesser}
       \begin{split}
       \frac{\mathrm{d}G^{<}(t,t')}{\mathrm{d}t} &= -i\left[h^{\mathrm{MF}}(t) G^{<}(t,t') + I^{<}(t,t')\right],\\
       I^{<}_{im}(t,t') &= -\sum_{klp}w_{iklp}(t) \Tilde{\mathcal{F}}_{lpmk}(t,t'),\\
        \frac{\mathrm{d}\Tilde{\mathcal{F}}_{lpmk}(t,t')}{\mathrm{d}t} &=  \sum_{qrsj} w_{qrsj}(t)\Big{[} G^{>,\mathrm{MF}}_{lq}(t)G^{>,\mathrm{MF}}_{pr}(t)G^{<,\mathrm{MF}}_{sk}(t) - G^{<,\mathrm{MF}}_{lq}(t)G^{<,\mathrm{MF}}_{pr}(t)G^{>,\mathrm{MF}}_{sk}(t)\Big{]} G^{<}_{jm}(t,t') \\&\hspace{15mm}-i\sum_{x}\Big{[} h^\mathrm{MF}_{lx}(t) \Tilde{\mathcal{F}}_{xpmk}(t,t') + h^\mathrm{MF}_{px}(t) \Tilde{\mathcal{F}}_{lxmk} (t,t') -\Tilde{\mathcal{F}}_{lpmx}(t,t') h^\mathrm{MF}_{xk}(t)\Big{]}.\\
   \end{split}
\end{equation}
The initial conditions cannot be fixed unambiguously as is the case for the retarded Green's function.  They now depend explicitly on the choice of diagonal.  In this paper we choose the following initial conditions,
\begin{equation}
\begin{split}
        G^{<}(t,t) &= G^{<,\mathrm{MF}}(t,t)\\
        \Tilde{\mathcal{F}}(t,t) &= 0
\end{split}
\end{equation}
Corresponding to an uncorrelated diagonal starting point.  In principle one could use the HF-GKBA collision integral to initialize $\Tilde{\mathcal{F}}(t,t')$ and still retain the overall numerical scaling of the scheme.
\newline
\newline
We stress that this scheme can be applied generally to any system defined by a set of time dependent single-particle basis states, $\{|\psi_i^{\mathrm{sp}}\rangle\}$,and a time-dependent single-particle static Hamiltonian $H^{0}(t)$,  which is the case in time-dependent density functional theory calculations\cite{ullrich_2012}. 
For this type of system we have the following definition of $G^{\mathrm{MF}}(t,t)$ and the time evolution operator, $\mathcal{U}(t,t')$
\begin{equation}
    \begin{split}
        G(t,t) &= \sum_{i}|\psi_i^\mathrm{sp(t)}\rangle\langle \psi_i^{\mathrm{sp}}(t)|\\
        \mathcal{U} &= \exp\left(-i{\int_{t}^{t'}\mathrm{d}\Bar{t}\medspace H^0(\bar{t})}\right)
    \end{split}
\end{equation}
This generality is due to the fact that the RT-DE depends only on the equations of motion in equation \eqref{eq:U_eom}, which are general for static Hamiltonians.  This is also true for the case of the $GW$ self-energy derived in section \ref{sec:GW_derivation} below.
\newline
\newline
\section{Derivation of the RT-DE Equations of Motion with the $GW$ self-energy}\label{sec:GW_derivation}
The $GW$ self-energy is expressed as
\begin{equation}
    \begin{split}
        \Sigma_{ij}^{\gtrless}(t,t') &= i\sum_{kl}W^\gtrless_{ilkj}(t,t')G^\gtrless_{kl}(t,t')\\
        W^\gtrless_{ilkj}(t,t') &= w_{ipkq}(t)\varepsilon_{pjql}^{-1,\gtrless}(t,t')
    \end{split}
\end{equation}
Following \cite{Joost_2020} we have the following explicit expression for $\varepsilon^{-1,\gtrless}(t,t')$,
\begin{equation}
    \begin{split}
        &\varepsilon_{ijkl}^{-1,\gtrless}(t,t') = -i\sum_{pq}w_{pjql}(t') G_{kp}^{\gtrless}(t,t')G_{qi}^\lessgtr(t',t) \\&- i\sum_{pqrs}w_{jrls}(t')\left[\int_{0}^t\mathrm{d}\bar{t}\left(G_{kp}^>(t,\bar{t})G_{qi}^<(\bar{t},t) - G_{kp}^<(t,\bar{t})G_{qi}^>(\bar{t},t) \right)\varepsilon^{-1,\lessgtr}(t',\Bar{t}) + \int_{0}^{t'}\mathrm{d}\bar{t}G_{kp}^\gtrless(t,\bar{t})G_{qi}^\lessgtr(\bar{t},t)\left(\varepsilon^{-1,>}_{rpsq} (t',\Bar{t}) - \varepsilon^{-1,<}_{rpsq} (t',\Bar{t})\right)\right]\\
&\implies\varepsilon_{ijkl}^{-1,\gtrless}(t,t) = -i\sum_{pq}w_{pjql}(t') G_{kp}^{\gtrless}(t)G_{qi}^\lessgtr(t) - i\sum_{pqrs}w_{jrls}(t)\int_{0}^t\mathrm{d}\Bar{t}\left(G^>_{kp}(t,\Bar{t})G^<_{qi}(\bar{t},t)\varepsilon^{-1,>}_{rpsq}(t,\bar{t}) - G^<_{kp}(t,\Bar{t})G^>_{qi}(\bar{t},t)\varepsilon^{-1,<}_{rpsq}(t,\bar{t})\right)\\
&= -i\sum_{pq}w_{pjql}(t') G_{kp}^{\gtrless}(t)G_{qi}^\lessgtr(t) + i\sum_{pqrs}w_{pjql}(t)\mathcal{G}_{kqip}(t).\\
\end{split}
\end{equation}

Where $\mathcal{G}(t)$ is a two-body correlator. Further, similarly to section \ref{sec:SB_derivation} we have the following definition of $\mathcal{F}(t,t')$.
\begin{equation}
    \begin{split}
        \mathcal{F}_{ijkl}(t,t') &= \sum_{pq}\int_{t'}^{t}d\Bar{t} \medspace\left[\varepsilon_{lpjq}^{-1,>}[G^{\mathrm{MF}}](t,\Bar{t})G^{>,\mathrm{MF}}_{ip}(t,\Bar{t}) - \varepsilon_{lpjq}^{-1,<}[G^{\mathrm{MF}}](t,\Bar{t})G^{<,\mathrm{MF}}_{ip}(t,\Bar{t})\right]G^{\mathrm{R}}_{qk}(\Bar{t},t')\\
    &= \sum_{pqr}\int_{t'}^{t}d\Bar{t} \medspace \mathcal{U}_{ir}(t,\Bar{t})\left[\varepsilon_{lpjq}^{-1,>}[G^{\mathrm{MF}}](t,\Bar{t})G^{>,\mathrm{MF}}_{rp}(\Bar{t},\Bar{t}) - \varepsilon_{lpjq}^{-1,<}[G^{\mathrm{MF}}](t,\Bar{t})G^{<,\mathrm{MF}}_{rp}(t,\Bar{t})\right]G^{\mathrm{R}}_{qk}(\Bar{t},t').
    \end{split}
\end{equation}
Where, as before, we make the approximation that the self-energy is evaluated with mean-field propagators.  Employing this approximation as well as equation \eqref{eq:U_eom} we can again derive an ODE scheme for $\mathcal{F}$ within the $GW$ self-energy.  The time-derivative of $\mathcal{F}$ is broken up as
\begin{equation}
    \frac{d}{dt}\mathcal{F}_{ijkl}(t,t') = \left[\frac{d}{dt}\mathcal{F}_{ijkl}(t,t')\right]_{\int} + \left[\frac{d}{dt}\mathcal{F}_{ijkl}(t,t')\right]_{\varepsilon} + \left[\frac{d}{dt}\mathcal{F}_{ijkl}(t,t')\right]_{\mathcal{U}}.
\end{equation}

\begin{equation}
    \begin{split}
        \left[\frac{d}{dt}\mathcal{F}_{ijkl}(t,t')\right]_{\int} &= \sum_{pqr}\mathcal{U}_{ir}(t,t)\left[\varepsilon_{lpjq}^{-1,>}[G^{\mathrm{MF}}](t,\Bar{t})G^{>,\mathrm{MF}}_{rp}(t) - \varepsilon_{lpjq}^{-1,<}[G^{\mathrm{MF}}](t,\Bar{t})G^{<,\mathrm{MF}}_{rp}(t)\right]G^{\mathrm{R}}_{qk}(t,t')\\
        &=\sum_{pq}\left[\varepsilon_{lpjq}^{-1,>}[G^{\mathrm{MF}}](t,t)G^{>,\mathrm{MF}}_{ip}(t) - \varepsilon_{lpjq}^{-1,<}[G^{\mathrm{MF}}](t,t)G^{<,\mathrm{MF}}_{ip}(t)\right]G^{\mathrm{R}}_{qk}(t,t')\\
        &=-i\sum_{pqxy}w_{xpyq}\bigg{[} \mathcal{G}_{jylx}[G^{\mathrm{MF}}](t)[G^{<,\mathrm{MF}}_{ip}(t) -G^{>,\mathrm{MF}}_{ip}(t)]G^{\mathrm{R}}_{qk}(t,t') \\ & \hspace{30mm}+ [G^{>,\mathrm{MF}}_{jx}(t)G^{<,\mathrm{MF}}_{yl}(t)G^{>,\mathrm{MF}}_{ip}(t) - G^{<,\mathrm{MF}}_{jx}(t)G^{>,\mathrm{MF}}_{yl}(t)G^{<,\mathrm{MF}}_{ip}(t)]G^{\mathrm{R}}_{qk}(t,t')\bigg{]}\\
    \end{split}
\end{equation}
The derivative with respect to $\mathcal{U}$ is as follows,

\begin{equation}
    \begin{split}
         \left[\frac{d}{dt}\mathcal{F}_{ijkl}(t,t')\right]_{\mathcal{U}} &= \sum_{pqr}\int_{t'}^{t}d\Bar{t} \medspace \frac{d}{dt}\mathcal{U}_{ir}(t,\Bar{t})\left[\varepsilon_{lpjq}^{-1,>}(t,\Bar{t})G^{>,\mathrm{MF}}_{rp}(t,\Bar{t}) - \varepsilon_{lpjq}^{-1,<}(t,\Bar{t})G^{<,\mathrm{MF}}_{rp}(t,\Bar{t})\right]G^{\mathrm{R}}_{qk}(\Bar{t},t')\\
         &=-i\sum_{pqrs}\int_{t'}^{t}d\Bar{t} \medspace h^{\mathrm{HF}}_{is}(t)\mathcal{U}_{sr}(t,\Bar{t})\left[\varepsilon_{lpjq}^{-1,>}(t,\Bar{t})G^{>,\mathrm{MF}}_{rp}(t,\Bar{t}) - \varepsilon_{lpjq}^{-1,<}(t,\Bar{t})G^{<,\mathrm{MF}}_{rp}(t,\Bar{t})\right]G^{\mathrm{R}}_{qk}(\Bar{t},t')\\
         &=-i\sum_{s} h^{\mathrm{HF}}_{is}(t)\mathcal{F}_{sjkl}(t,t')
    \end{split}
\end{equation}
For the derivative with respect to $\varepsilon^{-1,\gtrless}(t,t')$, by making the replacement $\varepsilon^{-1,\gtrless}(t,t') \rightarrow \varepsilon^{-1,\gtrless}[G^{\mathrm{MF}}](t,t')$, taking the time derivative and employing equation \eqref{eq:U_eom} we can derive,
\begin{equation}
    \begin{split}
        \frac{\mathrm{d}}{\mathrm{d}t}\varepsilon^{-1,\gtrless}(t,t') = -i\sum_{p}\Big{[}&h^{\mathrm{MF}}(t)\varepsilon_{ijpl}^{-1,\gtrless}(t,t') - \varepsilon_{pjkl}^{-1,\gtrless}(t,t')h^{\mathrm{MF}}(t)\Big{]} \\
        &- \sum_{pqrs}w_{prqs}(t)\left[G^{>,\mathrm{MF}}_{kp}(t)G^{<,\mathrm{MF}}_{qi}(t) - G^{<,\mathrm{MF}}_{kp}(t)G^{>,\mathrm{MF}}_{qi}(t)\right]\epsilon_{pjql}^{-1,\gtrless}(t,t')
    \end{split}
\end{equation}
\begin{equation}
    \begin{split}
        \left[\frac{d}{dt}\mathcal{F}_{ijkl}(t,t')\right]_{\varepsilon} &= \sum_{pqr}\int_{t'}^{t}d\Bar{t} \medspace \mathcal{U}_{ir}(t,\Bar{t})\left[\frac{d}{dt}\varepsilon_{lpjq}^{-1,>}(t,\Bar{t})G^{>,\mathrm{MF}}_{rp}(t,\Bar{t}) - \frac{d}{dt}\varepsilon_{lpjq}^{-1,<}(t,\Bar{t})G^{<,\mathrm{MF}}_{rp}(t,\Bar{t})\right]G^{\mathrm{R}}_{qk}(\Bar{t},t')\\
           &= -i\sum_{z}\left[ h^{\mathrm{HF}}_{jz}(t) \mathcal{F}_{izkl}(t,t') - \mathcal{F}_{ijkz}(t,t')h_{zl}^{\mathrm{HF}}(t)\right]\\&\hspace{20mm} - i\sum_{wxyz} w_{wxyz}[G_{jw}^{>,\mathrm{MF}}(t)G_{yl}^{<,\mathrm{MF}}(t) -  G_{jw}^{<,\mathrm{MF}}(t)G_{yl}^{>,\mathrm{MF}}(t)]\mathcal{F}_{izkx}(t,t')\\
\end{split}
\end{equation}
Combining the previous equations together we have the following expression for the ODE of $\mathcal{F}$
\begin{equation}\label{eq:GW_eom}
\begin{split}
         \frac{d}{dt}\mathcal{F}_{ijkl}(t,t') &=-i\sum_{pqxy}w_{xpyq}\bigg{[} \mathcal{G}_{jylx}[G^{\mathrm{MF}}](t)[G^{<,\mathrm{MF}}_{ip}(t) -G^{>,\mathrm{MF}}_{ip}(t)]G^{\mathrm{R}}_{qk}(t,t') \\ & \hspace{30mm}+ [G^{>,\mathrm{MF}}_{jx}(t)G^{<,\mathrm{MF}}_{yl}(t)G^{>,\mathrm{MF}}_{ip}(t) - G^{<,\mathrm{MF}}_{jx}(t)G^{>,\mathrm{MF}}_{yl}(t)G^{<,\mathrm{MF}}_{ip}(t)]G^{\mathrm{R}}_{qk}(t,t')\bigg{]}\\
         &\hspace{25mm}- i\sum_{wxyz} w_{wxyz}[G_{jw}^{>,\mathrm{MF}}(t)G_{yl}^{<,\mathrm{MF}}(t) -  G_{jw}^{<,\mathrm{MF}}(t)G_{yl}^{>,\mathrm{MF}}(t)]\mathcal{F}_{izkx}(t,t')\\
         &\hspace{25mm}-i\sum_{z}\left[ h^{\mathrm{HF}}_{iz}(t)\mathcal{F}_{zjkl}(t,t') + h^{\mathrm{HF}}_{jz}(t) \mathcal{F}_{izkl}(t,t') - \mathcal{F}_{ijkz}(t,t')h_{zl}^{\mathrm{HF}}(t)\right] \\
         \frac{d}{dt}G^{\mathrm{R}}_{ij}(t,t') &= -\sum_k h^{\mathrm{HF}}_{ik}(t)G^{\mathrm{R}}_{kj}(t,t') + \sum_{psq} w_{ipsq}(t) \mathcal{F}_{sqjp}(t,t').\\
         \end{split}
\end{equation}
Here, by replacing $G^{\mathrm{R}}(t,t')$ with $G^{\mathrm{<}}(t,t')$ we have the equations for the time off-diagonal lesser component.

Finally, we note that unlike the case of the second Born self-energy these equations of motion depend explicitly on a two particle correlation function evaluated with mean-field propagators, $\mathcal{G}_{ijkl}[G^{\mathrm{MF}}](t)$ where,
\begin{equation}
    \mathcal{G}_{ijkl}(t) = - \sum_{pq}\int_0^t d\bar{t}\medspace \left[\varepsilon_{lpjq}^{-1,>}[G^{\mathrm{MF}}](t,\bar{t})G^{>,\mathrm{MF}}_{ip}(t,\Bar{t})G_{qk}^{<,\mathrm{MF}}(\Bar{t},t) - \varepsilon_{lpjq}^{-1,<}[G^{\mathrm{MF}}](t,\bar{t})G^{<,\mathrm{MF}}_{ip}(t,\Bar{t})G_{qk}^{>,\mathrm{MF}}(\Bar{t},t)\right].
\end{equation}
Following similar steps as previously we arrive at the following equation,
\begin{equation}
    \begin{split}\label{eq:G2_eom}
        \frac{d}{dt}\mathcal{G}_{ijkl}(t) &= i \sum_{pqrs} w_{rpsq}\bigg{[}\left(G^{>,\mathrm{MF}}_{jr}(t)G^{<,\mathrm{MF}}_{sl}(t)G^{>,\mathrm{MF}}_{ip}(t)G_{qk}^<(t) - G^{<,\mathrm{MF}}_{jr}(t)G^{>,\mathrm{MF}}_{sl}(t)G^{<,\mathrm{MF}}_{ip}(t)G_{qk}^>(t)\right) \\&+ \mathcal{G}_{jslr}(t)\left(G^{<,\mathrm{MF}}_{ip}(t)G_{qk}^>(t) - G^{>,\mathrm{MF}}_{ip}(t)G_{qk}^<(t)\right)- G^{>,\mathrm{MF}}_{jr}(t)G^{<,\mathrm{MF}}_{sl}(t) - G^{<,\mathrm{MF}}_{jr}(t)G^{>,\mathrm{MF}}_{sl}(t)\mathcal{G}_{iqkp}(t)\bigg{]}\\
        &-i\sum_{p}\left[ h_{jp}^{\mathrm{MF}}(t)\mathcal{G}_{ipkl}(t)+h_{ip}^{\mathrm{MF}}(t)\mathcal{G}_{pjkl}(t) - h^{\mathrm{MF}}_{pl}(t)\mathcal{G}_{ijkp}(t) - h_{lp}^{\mathrm{MF}}(t)\mathcal{G}_{ijkp}(t)\right] 
    \end{split}
\end{equation}
Thus the RT-DE within the $GW$ self-energy approximation requires first the preparation of the mean-field Green's function, $G^{\mathrm{MF}}$, followed by the propagation of equation \eqref{eq:G2_eom}.  These inputs, along with initial conditions for $\mathcal{F}(t,t')$ can then be used to propagate equation \eqref{eq:GW_eom} within the chosen probe window.  The numerical scaling in the number of time steps remains effectively linear as with the second Born approximation.

\section{Equivalence to one shot correction}
Here we will show the equivalence of our scheme to a one-shot correction on top of a mean field reference Hamiltonian.  For a given self energy the Dyson equation for a one shot correction on top of a mean-field reference is given by
\begin{equation}\label{eq:dyson_eq}
    G(12) = G^{0}(12) + \int\int d3 d4  \medspace G^{0}(13)\Sigma[ G^{0}(34)] G(42)  
\end{equation}
The equation of motion that is being solved in the RT-DE is given by
\begin{equation}\label{eq:real_t_dyson}
    \left[i\partial_{t_1}- h^{0}[G(1)^{0}]\right]G^{\mathrm{R}}(12) - \int d3 \medspace \Sigma^{\mathrm{R}}[G^{0}(13)] G^{\mathrm{R}}(32) = \delta(12)
\end{equation}
The equation of motion for $G^{0}$ is given by
\begin{equation}
     \left[i\partial_{t_1}- h^{0}[G(1)^{0}]\right]G^{\mathrm{R,0}}(12) = \delta(12)
\end{equation}
From this we can define the left inverse of $G^{\mathrm{R,}0}(12)$ as $G^{-1}_\mathrm{R,0}(1)=\left[i\partial_{t_1}- h^{\mathrm{0}}[G(1)^{\mathrm{0}}]\right]$.  With this definition we can return to equation \eqref{eq:real_t_dyson} and rewrite it as,
\begin{equation}
    G^{-1}_\mathrm{R,0}(1)G^{\mathrm{R}}(12) - \int d3 \medspace \Sigma^{\mathrm{R}}[G^{\mathrm{0}}(13)] G^{\mathrm{R}}(32) = \delta(12).
\end{equation}
Multiplying both sides by $G^{\mathrm{R,0}}(31)$ and integrating over coordinate $1$ we have,
\begin{equation}
\begin{split}
       &\int d1 \medspace  G^{\mathrm{R,0}}(31)G^{-1}_\mathrm{R,0}(1)G^{\mathrm{R}}(12) - \int \int d1 d3 \medspace  G^{\mathrm{R,0}}(31)\Sigma^{\mathrm{R}}[G^{\mathrm{0}}(13)] G^{\mathrm{R}}(32) = \int d1 \medspace  G^{\mathrm{R,0}}(31)\delta(12),\\
     & \int d1 \medspace  \delta(13)G^{\mathrm{R}}(12) - \int \int d1 d3 \medspace  G^{\mathrm{R,0}}(31)\Sigma^{\mathrm{R}}[G^{\mathrm{0}}(13)] G^{\mathrm{R}}(32) = G^{\mathrm{R,0}}(32),\\
\end{split}
\end{equation}
Simplifying and rearranging we arrive at the final expression for the Dyson equation corresponding to the real time equation of motion in equation \eqref{eq:real_t_dyson}.  
\begin{equation}
    G^{\mathrm{R}}(32)  = G^{\mathrm{R,0}}(32) + \int \int d1 d3 \medspace  G^{\mathrm{R,0}}(31)\Sigma^{\mathrm{R}}[G^{\mathrm{0}}(13)] G^{\mathrm{R}}(32),\\
\end{equation}
Which is equivalent to (the retarded component of) equation \eqref{eq:dyson_eq}.

\section{Results for different non-equilibrium preperations}
In this section we will analyze additional results for the finite system.  In the following we provide comparison between exact diagonalization, the RT-DE, TD-HF and the HF-GKBA in the ground state as well as for several common non-equilibrium preperations.  The full Hamiltonian and different non-equilibrium preperations are given below.
\begin{equation}\label{eq:MB_ham}
\begin{split}
        \mathcal{H}  &= -J\sum_{\langle i,j\rangle}c^\dagger_ic_{j} + U\sum_{i}n_{i\uparrow}n_{i\downarrow} + \sum_{ij}h_{ij}^{\mathrm{N.E}}(t)c_{i}^\dagger c_j,\\
\end{split}
\end{equation} 
with the different non-equilibrium Hamiltonians given by
\begin{equation}\label{eq:h_NE}
     h_{ij}^{\mathrm{N.E}}(t) = \begin{cases}
       h^{\textrm{short wl}}_{ij}(t)&=\delta_{ij}E\cos\left(\frac{\pi r_i}{2}\right)\exp\left({-\frac{(t-t_0)^2}{2T_p^2}}\right)\\
       h^{\textrm{long wl}}_{ij}(t)&= \delta_{ij}Er_i\exp\left({-\frac{(t-t_0)^2}{2T_p^2}}\right)\\
       h^{\textrm{quench}}_{ij}(t)&= \delta_{ij}q\Theta(t)\Theta(N_q-i-1) 
    \end{cases}
\end{equation}
Here $r_i = \frac{N_s-1}{2} -i$ and $\Theta(t)$ is the Heaviside step function.  We refer to these as, a short wavelength excitation, a long wavelength excitation and a quench respectively.  We will now proceed by computing equation \eqref{eq:spectral_function} for the ground state of the system as well as for each of the non-equilibrium cases defined by equation \eqref{eq:h_NE}.

\begin{equation}\label{eq:spectral_function}
    \begin{split}
        \mathcal{A}(\omega,t_p) &= \int dt dt'\medspace \mathrm{e}^{-i\omega(t-t')}\mathcal{S}(t-t_p)\mathcal{S}(t'-t_p) \textrm{Tr}[G(t,t')],\\
        \mathcal{S}(t-t_p) &= \frac{1}{\delta\sqrt{2\pi}} \mathrm{e}^{-\frac{(t-t_p)^2}{2\delta^2}}.
    \end{split}
\end{equation}

\subsection{Ground state results}
\begin{figure}
    \centering
    \includegraphics[width=.7\textwidth]{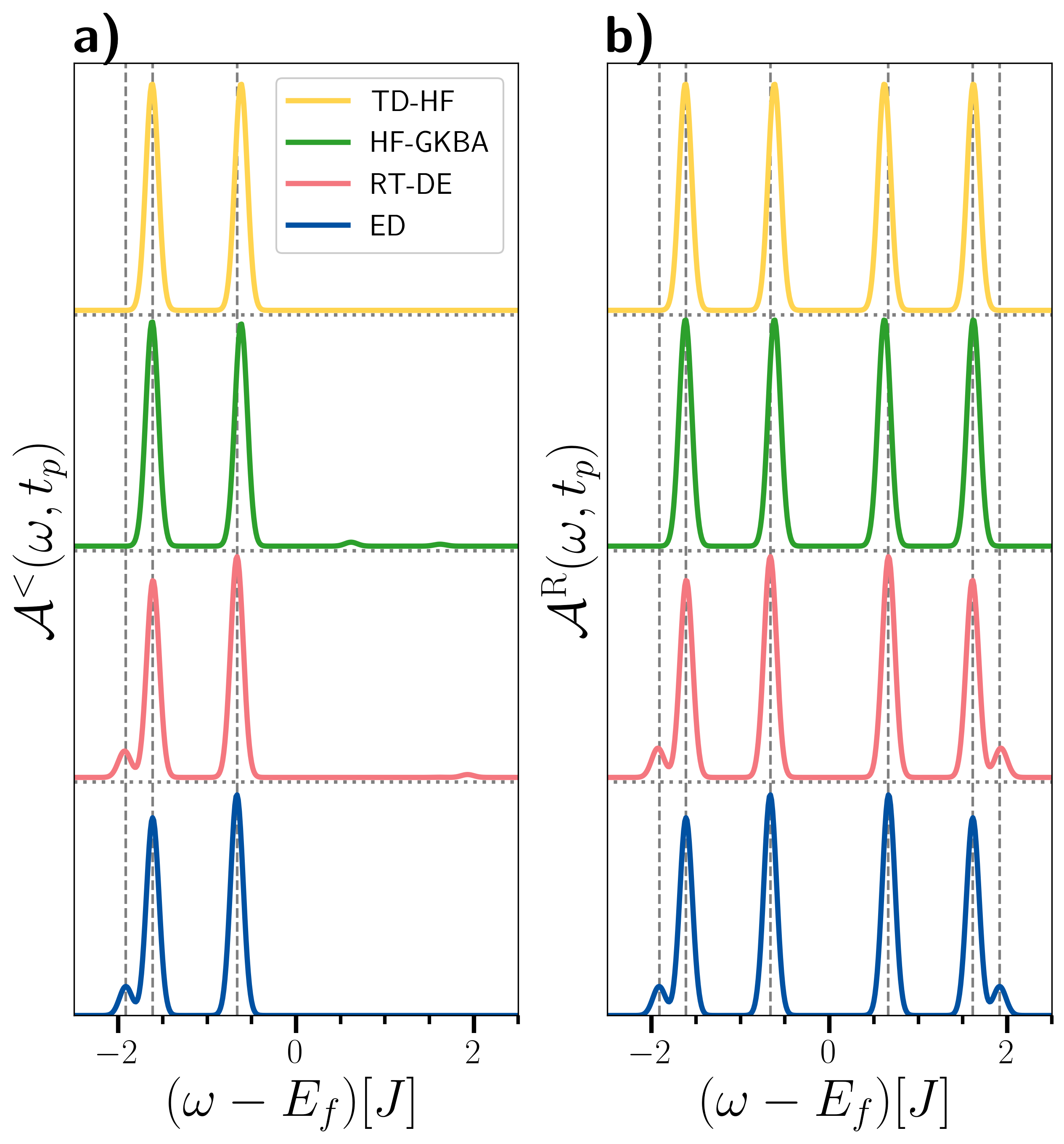}
    \caption{Ground state a) emission spectrum  and b) full spectrum found using  equation \eqref{eq:spectral_function} with $G = G^<$ and $G^{\mathrm{R}}$ respectively. Computed with exact diagonalization, Hartree-Fock, HF-GKBA and RT-DE in the model in equation \eqref{eq:MB_ham} with $h^{\mathrm{N.E}}(t) = 0$ and the following parameters: $U=1.0J$,  $N_s = 4$.}
    \label{fig:ground_state}
\end{figure}

We begin with the ground state properties of the system with $N_s= 4$ and $U = 1.0J$.  The spectral function of this system is shown in Fig. \ref{fig:ground_state} for each method. We use a Hartree-Fock calculation as the starting point for the RT-DE calculations.  We find this gives much better agreement than using the HF-GKBA as starting point. The reason for this will be discussed after we present and discuss the ground state results.

We observe excellent agreement between the RT-DE and exact diagonalization.  Significantly, the RT-DE picks up two important features that are missed by the other approximate methods.  Firstly, the shift in the peaks nearest the Fermi energy and secondly the appearance of satellite peaks near $\omega\approx \pm 1.9J$.  These features can be attributed to the static and dynamical parts of the collision integral respectively. Due to the neglect of a collision integral both features are missed entirely by the HF-GKBA and the Hartree-Fock calculation.   

Another significant observation is that the HF-GKBA offers no improvement on the spectral properties produced by Hartree-Fock.  Furthermore, in Fig. \ref{fig:ground_state} a) we see the appearance of two small peaks above the Fermi energy, at $\omega\approx 0.6J$ and $\omega \approx 1.6J$.  Since this is a ground state calculation there should only be peaks present below the Fermi energy.  There is a simple explanation for this observation. The HF-GKBA ground state is prepared by adiabatic switching and does not correspond to the Hartree-Fock ground state. This necessarily means that at the Hartree-Fock level it must be an excited state.  Hence we see the appearance of peaks above the Fermi energy.  These are artificial excited states marks an inconsistency present in the HF-GKBA.  This result also suggests that using Hartree-Fock is a better starting point for the RT-DE compared to the HF-GKBA.  Simply put the Hartree-Fock spectral function is closer to the true spectral function.  Thinking of this as adding a correction upon a given theory it is only logical that the closer our starting point is to the true result the more effective the correction will be.  The RT-DE also shows a small artificial excited state for $\mathcal{A}^<(\omega,t_p)$ in Fig. \ref{fig:ground_state} a) around $\omega \approx 1.9J$ which we attribute to the choice of initial condition for $G^<(t,t')$ on the diagonal.  We are currently further investigating the choice of this initial condition.

The latter indicates that the dynamics is captured well, while the inconsistency observed in the lesser component of the spectral function is a result the particular choice of boundary conditions of the  lesser component of the GF, which will require further investigation and optimization. Yet, our results demonstrate that  the quantities such as peak positions, band gap and the band-edge state energy renormalization, exciton energies, etc., are accurately obtained as they are extracted from the retarded component of the spectral function which is independent of the choice for $G^<$. 
\subsection{Short wavelength excitation}
\begin{figure}
    \centering
    \includegraphics[width=.7\textwidth]{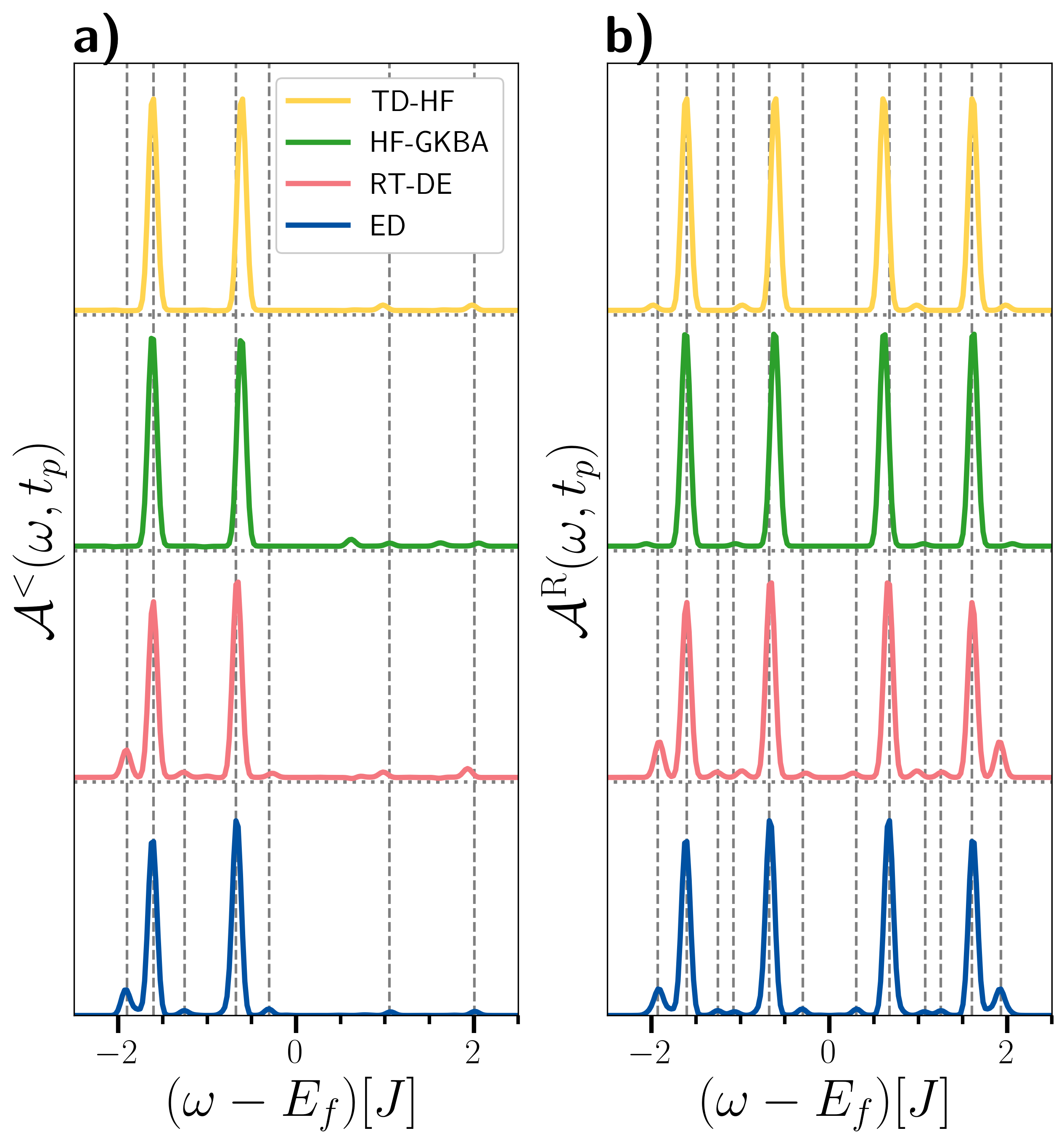}
    \caption{Time resolved a) emission spectrum  and b) full spectrum  computed with exact diagonalization, RT-DE, TD-HF and the HF-GKBA. Computed for the model in equation \eqref{eq:MB_ham} with $h^{\mathrm{N.E}} = h^{\textrm{Short wl}}$ and the following parameters: $U=1.0J$,  $N_s = 4$, $E = 0.5J$, $t_0 = 5J^{-1}$ and $T_p = 0.5J^{-1}$ after time evolving to $T_{\mathrm{max}} = 200J^{-1}$.  The probe width is taken to be $\delta = 15$ and $t_p = 100J^{-1}$.}
    \label{fig:short_wl}
\end{figure}

Next, in Fig. \ref{fig:short_wl}, we show non-equilibrium results for the previous system with $h^{\mathrm{N.E}}(t) = h^{\textrm{Short wl}}(t)$,  $E = 0.5J, t_0 = 5J^{-1}$ and $T_p = 0.5J^{-1}$.  The lesser component has been discussed in the main text but here we show both the emission and full spectrum of the system.  We note that for the lesser component the RT-DE has additional peaks at $\omega \approx 0.8J$ and $\omega \approx -1.0J$ while the retarded component has the correct number of peaks and are in excellent agreement with the exact peak positions. This shows that despite the need to further investigate the starting point for the lesser component of the spectrum, the retarded spectrum can still give accurate information about quantities such as peak positions, band gaps and ionization potentials.  

\subsection{Long wavelength excitation}
\begin{figure}
    \centering
    \includegraphics[width=.7\textwidth]{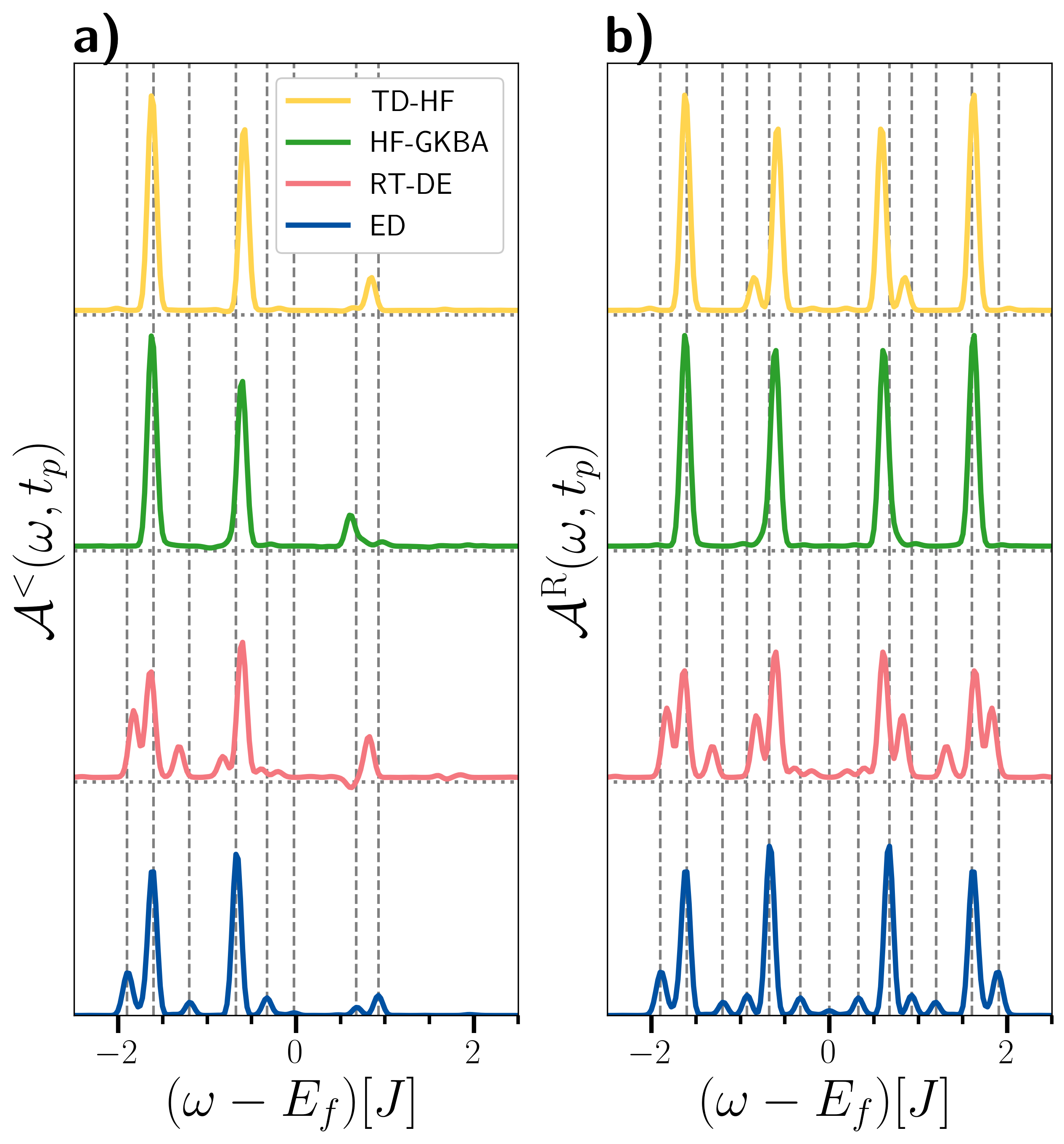}
    \caption{Time resolved a) emission spectrum  and b) full spectrum  computed with exact diagonalization, RT-DE, TD-HF and the HF-GKBA. Computed for the model in equation \eqref{eq:MB_ham} with $h^{\mathrm{N.E}} = h^{\textrm{Long wl}}$ and the following parameters: $U=1.0J$,  $N_s = 4$, $E = 0.5J$, $t_0 = 5J^{-1}$ and $T_p = 0.5J^{-1}$ after time evolving to $T_{\mathrm{max}} = 200J^{-1}$.  The probe width is taken to be $\delta = 15$ and $t_p = 100J^{-1}$.}
    \label{fig:long_wl}
\end{figure}

Fig. \ref{fig:long_wl} shows the same time resolved spectra as the previous sections, now for the case of $h^{\mathrm{N.E}}(t) = h^{\textrm{Long wl}}(t)$ with $E = 0.5J$, $t_0 = 5J^{-1}$ and $T_p = 0.5J^{-1}$.  Compared to the short wavelength excitation the deviation from the ground state spectrum is stronger in the case of the long wavelength excitation.  This is reflected in the quality of the approximate results.  

We see TD-HF and the HF-GKBA perform quite poorly for both the lesser and retarded component of the time resolved spectral function.  For the lesser component both the HF-GKBA pick up the peaks at $\omega\approx-1.6J$, $\omega\approx -0.7J$, $\omega\approx0.7J$ and $\omega \approx0.9J$,   however the HF-GKBA severely over estimates the peak magnitude at $\omega\approx0.7J$ while severely underestimating the magnitude at $\omega\approx0.9J$.  This also leads to the HF-GKBA well underestimating the peaks at $\omega\approx \pm0.9J$ in the retarded component in Fig. \ref{fig:long_wl} b).  TD-HF also shows peaks at $\omega\approx-1.6J$, $\omega\approx -0.7J$, $\omega\approx0.7J$ and $\omega \approx0.9J$, and the relative heights of the peaks are captured much more closely here than with the HF-GKBA.  This is reflected in the ability of TD-HF to capture the peaks around $\omega \approx 1.9J$.  Again we see the spectral results produced by the HF-GKBA are worse than those produced by TD-HF.  Regardless of their relative quality, both HF-GKBA and TD-HF miss several peaks arising from dynamical correlations, an error that is improved by using the RT-DE.

Despite the stronger deviation from the ground state the RT-DE still performs excellently compared to the exact diagonalization results. For the lesser component the peaks at $\omega\approx-1.6J$ and $\omega\approx-0.7J$ are captured very close. Similarly the peak at $\omega \approx 0.9J$ is captured as well by the RT-DE and is nearly identical to the TD-HF result, while the peak at $\omega \approx 0.7J$ has negative spectral weight.  Interestingly despite having this negative spectral weight, the relative height of the two peaks at $\omega\approx 0.7J$ and $\omega\approx 0.9J$ is closer to the exact result than for the other approximate methods. The reason for this negative weight requires further investigation but is likely due to the strength of the perturbation from equilibrium or the inexact initial condition for $G^<(t,t')$.  The peaks at $\omega \approx -1.9J$, $\omega \approx -1.2J$ and $\omega \approx -0.4J$ are also well captured by the RT-DE.  Significantly, these peaks are missed entirely by the mean-field methods. The lesser component shows an additional peak at $\omega \approx -0.9J$ which again we attribute to the initial condition for $G^<(t,t')$.  

Turning to the retarded component in Fig. \ref{fig:long_wl} b) we see the RT-DE is in excellent agreement with the exact result and improves significantly on the HF-GKBA and TD-HF result.  However, we note that the exact result shows a density of states at the Fermi level whereas the RT-DE has split this density to form two peaks on either side of the Fermi level at $\omega \approx \pm .3J$.
\subsection{System quench }
\begin{figure}
    \centering
    \includegraphics[width=.7\textwidth]{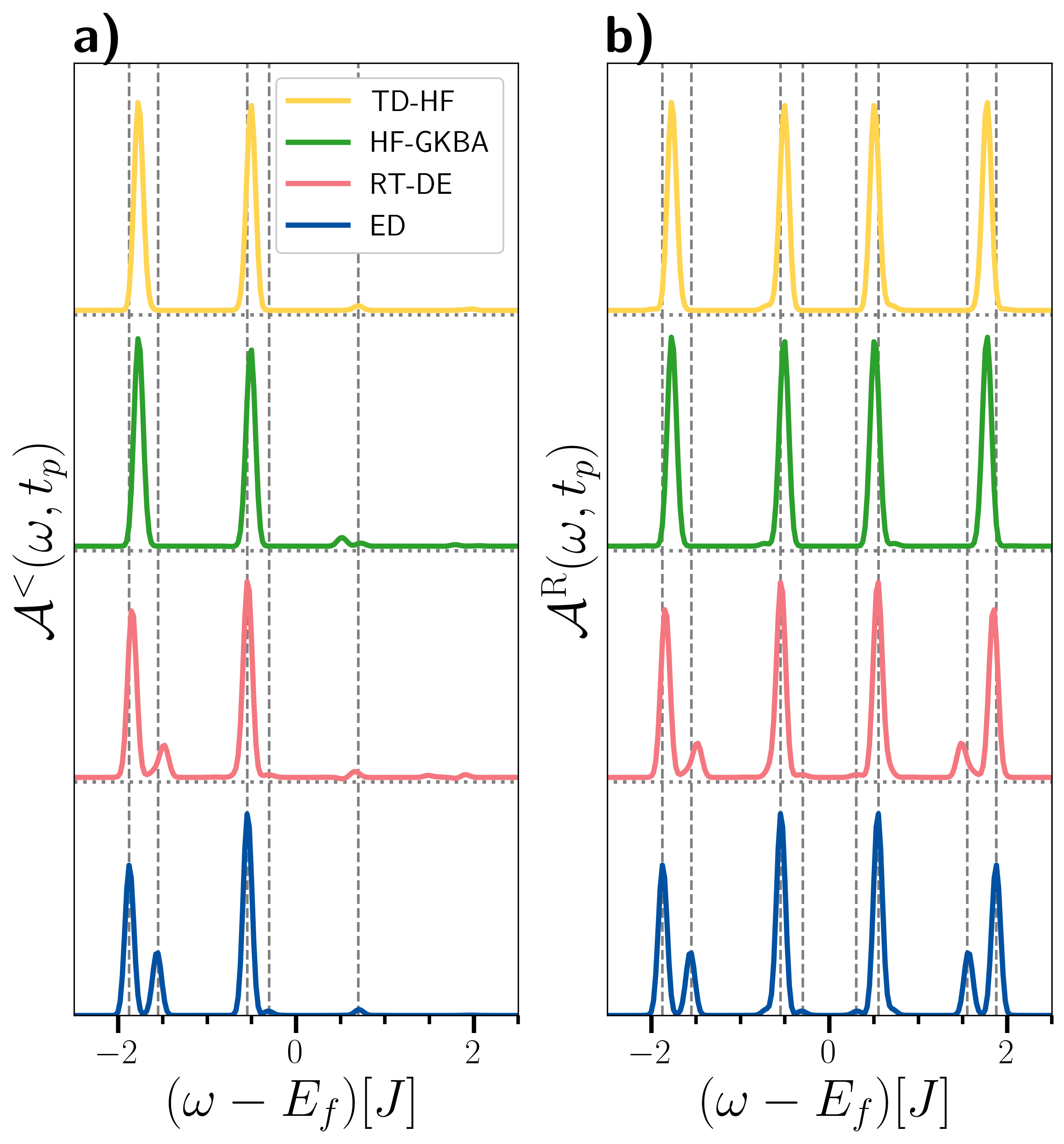}
    \caption{Time resolved a) emission spectrum  and b) full spectrum  computed with exact diagonalization, RT-DE, TD-HF and the HF-GKBA. Computed for the model in equation \eqref{eq:MB_ham} with $h^{\mathrm{N.E}} = h^{\textrm{quench}}$ and the following parameters: $U=1.0J$,  $N_s = 4$, $N_q=2$, and $q=1.0$ after time evolving to $T_{\mathrm{max}} = 200J^{-1}$.  The probe width is taken to be $\delta = 15$ and $t_p = 100J^{-1}$.}
    \label{fig:quench}
\end{figure}

Finally, we discuss the results in Fig. \ref{fig:quench} . Here we show the spectrum for $h^{\mathrm{N.E}}(t) = h^{\textrm{quench}}(t)$ with $N_q = 2$ and $q = 1.0J$.  For the lesser component all three methods capture the peaks at $\omega\approx-1.9J$, $\omega\approx-0.5J$ and $\omega \approx 0.6J$. The RT-DE has better agreement with the main peak positions at $\omega\approx-1.9J$, $\omega\approx-0.5J$, than the HF-GKBA and TD-HF. The HF-GKBA has an additional larger peak around $\omega \approx 0.5J$ which is due to the inconsistency between the HF-GKBA density matrix and the TD-HF equation of motion.  This matter is described in more detail in the discussion of the ground state spectrum. 
The RT-DE also has a very small additional peak around $\omega \approx1.9J$, again we attribute this to the choice of initial condition for $G^<(t,t')$.  As in the previous examples we again see the RT-DE picks up features that are completely missed by the mean-field approaches.  In this case we see  peaks at $\omega \approx -1.6J$ and $\omega \approx -0.3J$.  The RT-DE not only predicts the existence of these additional peaks but also match their positions very accurately.  For the retarded component we see similarly excellent agreement between the RT-DE and the exact result as has been observed in the previous sections

\FloatBarrier
\section{Comparison of Diagonal time evolution for Hubbard model}
\begin{figure}
    \centering
    \includegraphics[width=\textwidth]{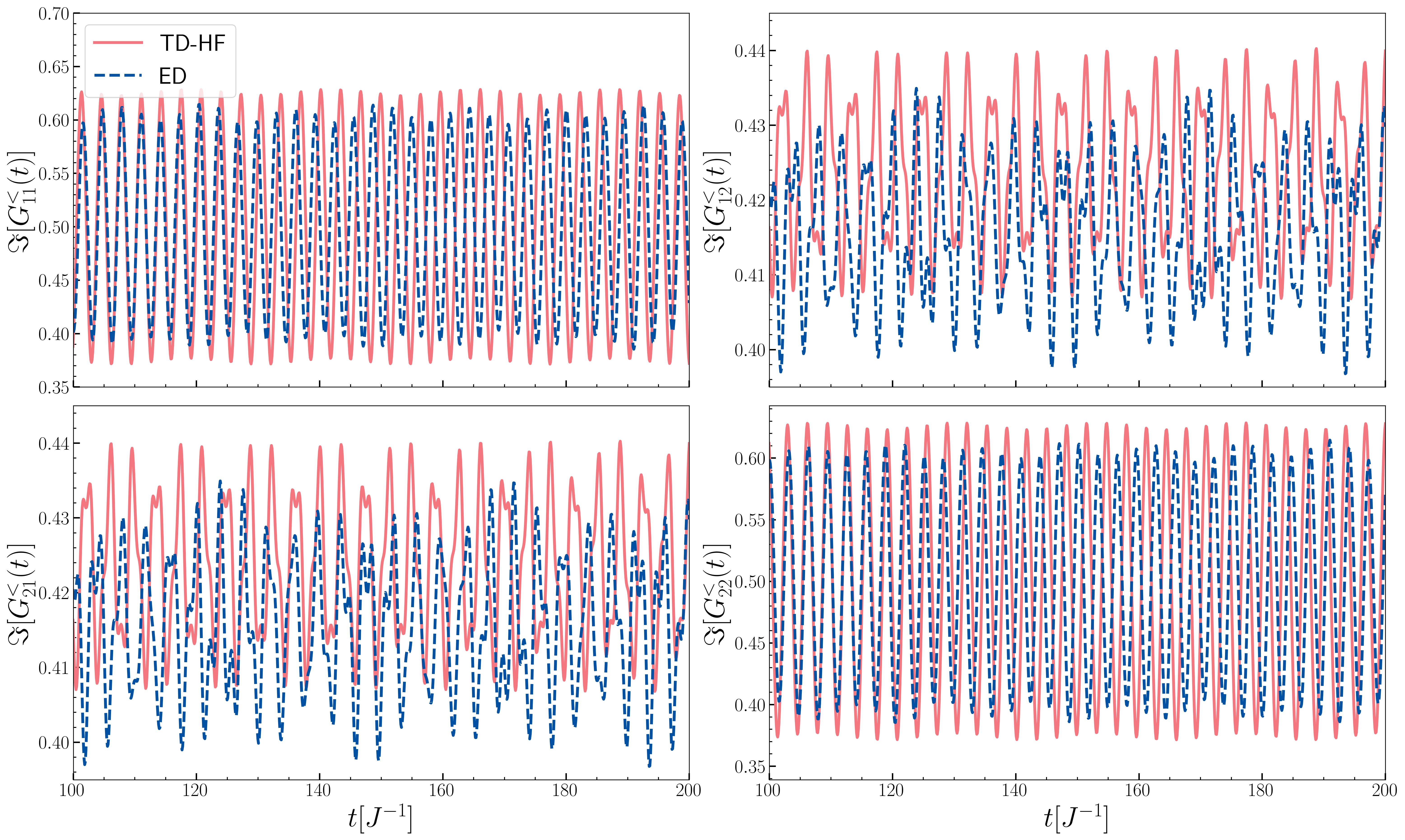}
    \caption{Comparison between exact diagonalization and time-dependent Hartree-Fock for the non-equilibrium dynamics of a selection of the imaginary components of the time diagonal Green's function. Computed for the model in equation \eqref{eq:MB_ham} with $h^{\mathrm{N.E}}(t) = h^{\textrm{Short wl}}(t)$ and the following parameters: $U=1.0J$,  $N_s = 4$, $E = 0.5J$, $t_0 = 5J^{-1}$ and $T_p = 0.5J^{-1}$ after time evolving to $T_{\mathrm{max}} = 200J^{-1}$.  The probe width is taken to be $\delta = 15$ and $t_p = 150J^{-1}$.}
    \label{fig:diag_comparision_swl}
\end{figure}

\begin{figure}
    \centering
    \includegraphics[width=\textwidth]{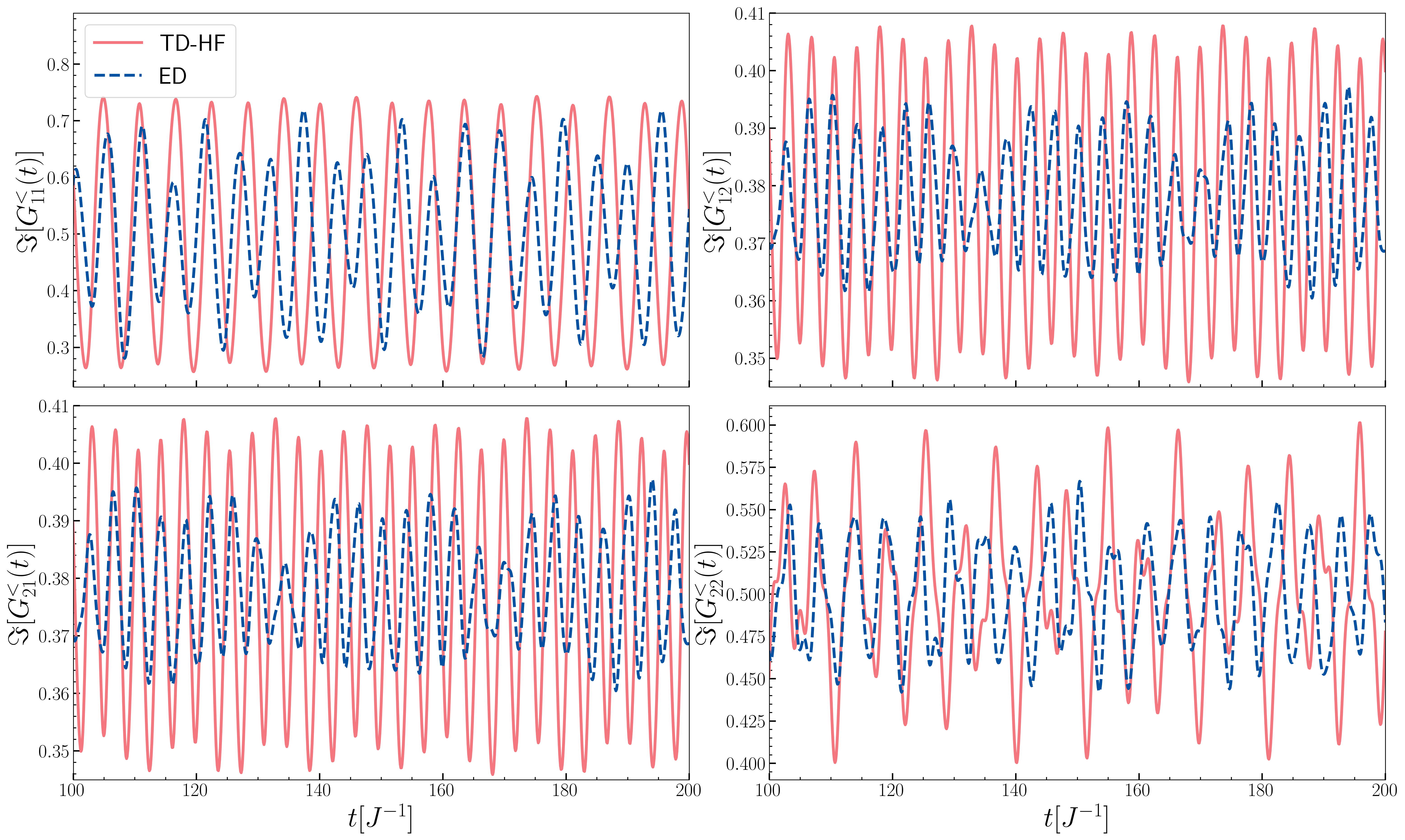}
    \caption{Comparison between exact diagonalization and time-dependent Hartree-Fock for the non-equilibrium dynamics of a selection of the imaginary components of the time diagonal Green's function. Computed for the model in equation \eqref{eq:MB_ham} with $h^{\mathrm{N.E}}(t) = h^{\textrm{Long wl}}(t)$ and the following parameters: $U=1.0J$,  $N_s = 4$, $E = 0.5J$, $t_0 = 5J^{-1}$ and $T_p = 0.5J^{-1}$ after time evolving to $T_{\mathrm{max}} = 200J^{-1}$.  The probe width is taken to be $\delta = 15$ and $t_p = 150J^{-1}$.}
    \label{fig:diag_comparision_lwl}
\end{figure}

\begin{figure}
    \centering
    \includegraphics[width=\textwidth]{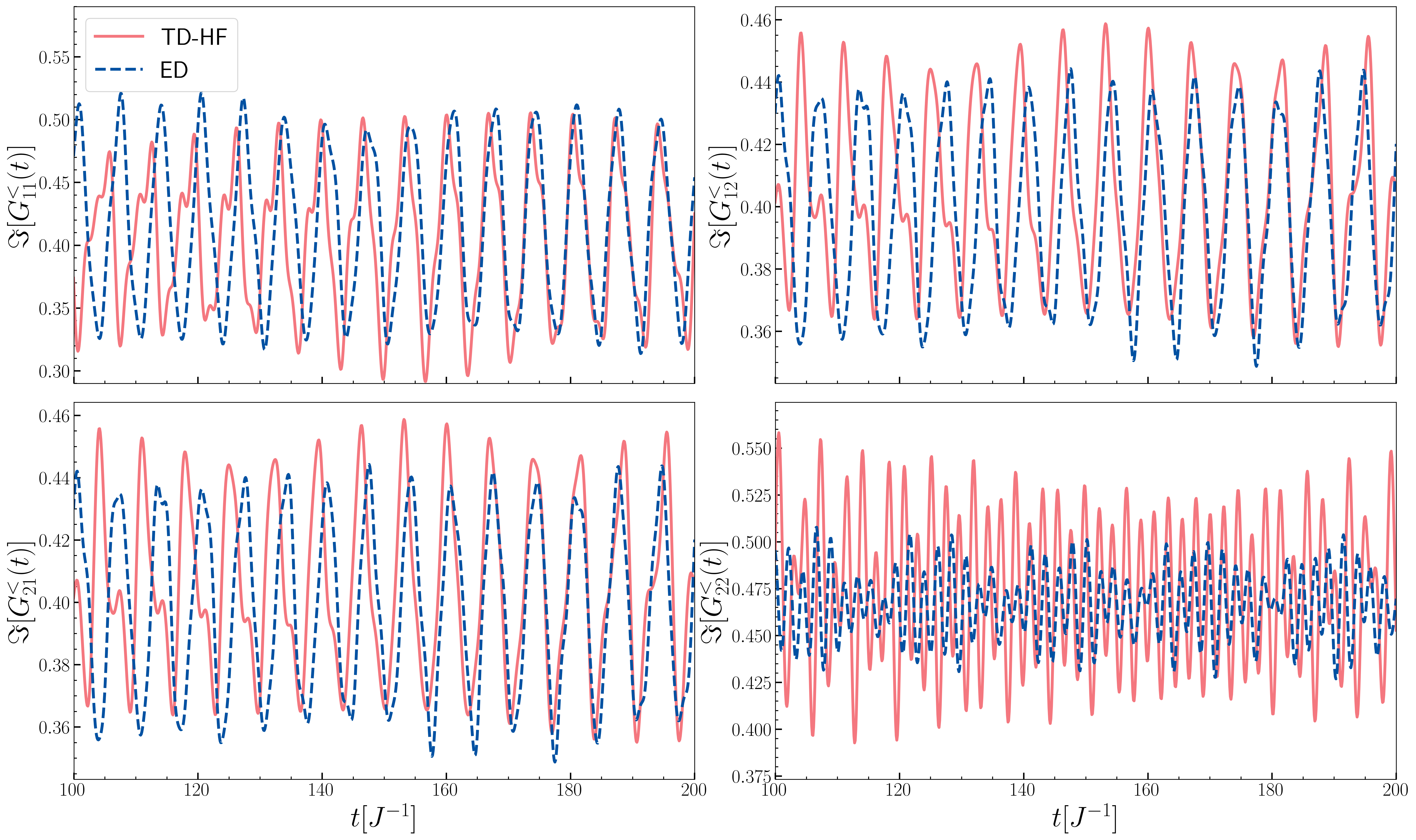}
    \caption{Comparison between exact diagonalization and time-dependent Hartree-Fock for the non-equilibrium dynamics of a selection of the imaginary components of the time diagonal Green's function. Computed for the model in equation \eqref{eq:MB_ham} with $h^{\mathrm{N.E}}(t) = h^{\textrm{quench}}(t)$ and the following parameters: $U=1.0J$,  $N_s = 4$, $N_q = 2$ and $q = 1.0J$ after time evolving to $T_{\mathrm{max}} = 200J^{-1}$.  The probe width is taken to be $\delta = 15$ and $t_p = 150J^{-1}$.}
    \label{fig:diag_comparision_quench}
\end{figure}

In this section we show a comparison between the results produced by TD-HF and exact diagonalization for the model each of the non-equilibrium Hamiltonians in equation \eqref{eq:h_NE}.  The purpose of this comparison is to see how well the time-diagonal component is captured by TD-HF.  In Fig. \ref{fig:diag_comparision_swl}-\ref{fig:diag_comparision_quench} we show a selection of imaginary components of the Green's function computed with TD-HF and exact diagonalization within the probe window.     In all three cases the TD-HF result generally performs well compared to the exact diagonalization result.  The primary difference is the magnitude of oscillation of the TD-HF result is larger than that of the exact result, however the main frequencies of oscillation are captured well.

The closeness between result of the dynamics along the diagonal suggests that TD-HF is a good initial guess for our off-diagonal reconstruction.  When going to more strongly correlated systems it may be necessary to use a more sophisticated mean-field Hamiltonian along the diagonal.  
\FloatBarrier
\FloatBarrier
\FloatBarrier
\FloatBarrier\FloatBarrier
\FloatBarrier
\section{Implementation details}
For the time evolution's shown in this manuscript we use 4th order Runge-Kutte algorithm with a time step of $dt = 0.05J^{-1}$. For the integration in equation \eqref{eq:spectral_function} we perform numerical integration using the rectangle rule with the same time step.  The frequency spacing used to compute $\mathcal{A}(\omega,t_p)$ is $d\omega = .05J$.  The Fourier transform of the spectral function in Fig. 3 of the main text was performed with fftw.

\bibliography{Bib_SI}